\documentclass{emulateapj}
\usepackage{amsmath}

\newcommand{\mh}{{M_\bullet}}
\newcommand{\rh}{r_\mathrm{infl}}

\shorttitle{Orbital structure of triaxial nuclei}
\shortauthors{Merritt and Vasiliev}

\begin{document}

\title{Orbits Around Black Holes in Triaxial Nuclei}

\author{David Merritt}
\affil{Department of Physics and Center for Computational Relativity
and Gravitation, Rochester Institute of Technology, Rochester, NY 14623}
\email{merritt@astro.rit.edu}

\author{Eugene Vasiliev}
\affil{Lebedev Physical Institute, Leninsky prospekt 53, Moscow, Russia}
\email{eugvas@lpi.ru}

\begin{abstract}
We discuss the properties of orbits within the influence sphere 
of a supermassive black hole (BH), in the case that the surrounding 
star cluster is nonaxisymmetric.  
There are four major orbit families; one of these, the pyramid orbits, 
have the interesting property that they can approach arbitrarily closely 
to the BH.  
We derive the orbit-averaged equations of motion and show that 
in the limit of weak triaxiality, the pyramid orbits are integrable: 
the motion consists of a two-dimensional libration of the major axis 
of the orbit about the short axis of the triaxial figure, 
with eccentricity varying as a function of the two orientation angles, 
and reaching unity at the corners.  
Because pyramid orbits occupy the lowest angular momentum regions of 
phase space, they compete with collisional loss cone repopulation 
and with resonant relaxation in supplying matter to BHs. 
General relativistic advance of the periapse 
dominates the precession for sufficiently eccentric orbits, and 
we show that relativity imposes an upper limit to the eccentricity: 
roughly the value at which the 
relativistic precession time is equal to the time for torques to 
change the angular momentum.
We argue that this upper limit to the eccentricity
should apply also to evolution driven by resonant relaxation,
with potentially important consequences for the rate of
extreme-mass-ratio inspirals in low-luminosity galaxies.
In giant galaxies, we show that capture of stars on pyramid orbits 
can dominate the feeding of BHs, 
at least until such a time as the pyramid 
orbits are depleted; however this time can be of order a Hubble time.
\end{abstract}


\section{Introduction}

Following the demonstration that self-consistent equilibria 
could be constructed for triaxial galaxy models
\citep{Schwarzschild1979,Schwarzschild1982},
observational evidence gradually accumulated for non-axisymmetry
on large (kiloparsec) scales in early-type galaxies 
\citep{Franx1991,Statler2004,Cappellari2007}.
On smaller scales,
imaging of the centers of galaxies also revealed a wealth of 
features in the stellar distribution that are not consistent with axisymmetry,
including bars, bars-within-bars, and nuclear spirals
\citep{Shaw1993,ErwinSparke2002,Seth2008}.
In the nuclei of low-luminosity galaxies, the non-axisymmetric features
may be recent or recurring, associated with ongoing star formation;
in luminous elliptical galaxies, central relaxation times are so
long that triaxiality, once present, could persist for the age
of the universe.

In a triaxial nucleus, torques from the stellar potential 
can induce gradual changes in the eccentricities of stellar orbits,
allowing stars to find their way into the central BH.
Gravitational two-body scattering also drives stars into the central BH,
but only on a time scale of order the central relaxation time,
which can be very long, particularly in the most luminous galaxies.
Simple arguments suggest that the feeding of stars to the central
BHs in many galaxies is likely to be dominated by large-scale
torques rather than by two-body relaxation \citep[e.g.][]{MerrittPoon2004}.

This paper discusses the character of orbits near a
supermassive BH in a triaxial nucleus.
The emphasis is on low-angular-momentum, or ``centrophilic,'' orbits, 
the orbits that come closest to the BH.
Self-consistent modelling \citep{PoonMerritt2004}
reveals that a large fraction of the orbits in triaxial
BH nuclei can be centrophilic.

Within the BH influence sphere, orbits are nearly Keplerian,
and the force from the distributed mass can be treated as a small 
perturbation which causes the orbital elements (inclination, eccentricity)
to change gradually with time.
A standard way to deal with such motion is via orbit averaging
\cite[e.g.][]{SandersVerhulst1985}, i.e., 
averaging the equations of motion over the
short time scale associated with the unperturbed Keplerian motion.
The result is a set of equations describing the slow evolution
of the remaining orbital elements due to the perturbing forces.
This approach was followed by \cite{SridharTouma1999} for motion
in an axially-symmetric nucleus containing a massive BH, 
and by \cite{SambhusSridhar2000} for 
motion in a constant-density triaxial nucleus.

In their discussion of motion in triaxial nuclei, 
\citet{SambhusSridhar2000} passed over one important class of orbit:
the centrophilic orbits, i.e., orbits that pass arbitrarily close to the BH.
Examples of centrophiliic orbits include the two-dimensional
``lens'' orbits \citep{SridharTouma1997,SridharTouma1999} and
the three-dimensional ``pyramids'' \citep{MerrittValluri1999, PoonMerritt2001}.
Centrophilic orbits are expected to dominate
the supply of stars and stellar remnants to a supermassive BH
\citep[e.g.][]{MerrittPoon2004}
and are the focus of the current paper.

The paper is organized as follows. 
In \S\ref{sec_potential} we present a model for the gravitational potential 
of a triaxial nuclear star cluster, which is more general than that 
studied in \cite{SambhusSridhar2000}, but which has many of the same dynamical 
features. 
Then in \S\ref{sec_orbitavg} we write down the orbit-averaged 
equations of motion, and in \S\ref{sec_analysis} present a detailed 
analytical study of their solutions, with emphasis on the case where the 
triaxiality is weak and the eccentricity large.
In this limiting case, the averaged equations of motion turn out to be fully
integrable.
In \S\ref{sec_capture} we derive the equations that describe 
the rate of capture of stars on pyramid orbits by the BH. 
Comparison of orbit-averaged treatment with real-space motion is made in 
\S\ref{sec_realspace}, to test the applicability of the former. 
In \S\ref{sec_GR} we consider the effect of general relativity 
on the motion, which imposes an effective upper limit on the eccentricty.
\S\ref{sec_RR} discusses the connection with resonant relaxation:
we argue that a similar upper limit to the eccentricity should
characterize orbital evolution in the case of resonant relaxation.
Finally, in \S\ref{sec_estimates} we make some quantitative 
estimates of the importance of pyramid orbits 
for capture of stars in galactic nuclei.
\S\ref{sec_summary} sums up.

\section{Model for the nuclear star cluster}  \label{sec_potential}

Consider a nucleus consisting of a BH, a spherical star cluster,
and an additional triaxial component.
An expression for the gravitational potential that includes
the three components is
\begin{eqnarray} 
\Phi({\bf r}) &=& -\frac{G\mh}{r} +
\Phi_s\left(\frac{r}{r_0}\right)^{2-\gamma} \\
  &+& 2\pi\, G\rho_t\,\left(T_x x^2 + T_y y^2 + T_z z^2\right). \nonumber
\label{Equation:TriaxPotential}
\end{eqnarray}
The second term on the right hand side is the potential
of a spherical star cluster with density 
$\rho(r) = \rho_s (r/r_0)^{-\gamma}$; the coefficient $\Phi_s$ is
given by
$$
\Phi_s = \frac{4\pi G}{(3-\gamma)(2-\gamma)} \rho_s r_0^2.
$$
The scale radius $r_0$ may be chosen arbitrarily but it is convenient 
to set $r_0=\rh$, with $\rh$ the radius at which the enclosed
stellar mass is twice $\mh$:
\begin{equation}\label{eq:defrh}
\rh = \left(\frac{3-\gamma}{2\pi} \frac{\mh}{\rho_s}\right)^{1/3}.
\end{equation}
The third term is the potential of a homogeneous
triaxial ellipsoid of density $\rho_t$;
this term can also be interpreted as a first approximation to the 
potential of a more general, inhomogeneous triaxial component.
In the former case, the dimensionless coefficients ($T_x,T_y,T_z$) 
are expressible in terms of the
axis ratios $(p,q)$ of the ellipsoid via elliptic integrals
\citep{ChandraEllipsoids}.
The $x(z)$ axes are assumed to be the
long(short) axes of the triaxial figure;
this implies $T_x\le T_y\le T_z$.
In what follows we will generally assume $\rho_t\ll\rho_s(r_0)$,
i.e. that the triaxial bulge has a low density compared with
that of the spherical cusp at $r=\rh$.

\section{Orbit-averaged equations}  \label{sec_orbitavg}

Within the BH influence sphere, orbits are nearly 
Keplerian\footnote{We consider general relativistic corrections in 
\S\ref{sec_GR}.}
and the force from the distributed mass can be treated as a small 
perturbation which causes the elements of the orbit
(inclination, eccentricity etc.) to change gradually with time.
A standard way to deal with such motion 
\citep[e.g.][]{SandersVerhulst1985} is to
average the equations over the coordinate executing
the most rapid variation, e.g., the radius.
The result is a set of equations describing the slow evolution
of the remaining variables due to the perturbing forces.

We begin by transforming from Cartesian coordinates to 
action-angle variables in the Kepler problem.
Following \citet{SridharTouma1999} and \citet{SambhusSridhar2000},
we adopt the Delaunay variables \citep[e.g.][]{GoldsteinMechanics}
to describe the unperturbed motion.

Let $a$ be the semi-major axis of the Keplerian orbit.
The Delaunay action variables are 
the radial action $I=(G\mh a)^{1/2}$, 
the angular momentum $L$, and the projection of $\mathbf{L}$ onto
the $z$ axis $L_z$.
The conjugate angle variables are the mean anomaly $w$,
the argument of the periapse $\varpi$, 
and the longitude of the ascending node $\Omega$.
In the Keplerian case, five of these are constants;
the exception is $w$ which increases linearly with time at a rate
\begin{equation}  
\nu_r=(GM_\bullet)^2/I^3
\label{Equation:nu_r}
\end{equation}

In terms of the new variables, the Hamiltonian is
\begin{equation}
{\cal H} = -\frac{1}{2}\left(\frac{GM_\bullet}{I}\right)^2 + 
\Phi_p(I,L,L_z,w,\varpi,\Omega);
\label{Equation:Hamiltonian}
\end{equation}
the first term is the Keplerian contribution and $\Phi_p$,
the ``perturbing potential'', contains the contributions from
the spherical and triaxial components of the distributed mass.
This transformation is completely general if we interpret the new variables 
as instantaneous (osculating) orbital elements.
However if we assume that the perturbing potential is small compared
with the point-mass potential, 
the rates of change of these variables 
(again with the exception of $w$) will be small compared with 
the radial frequency $\nu_r$, and
the new variables can be regarded as approximate orbital elements 
that change little over a radial period $P\equiv 2\pi/\nu_r$.
Accordingly, we average the Hamiltonian over the fast angle $w$:
\begin{subequations}
\begin{eqnarray}
\overline{\cal H} &=& -\frac{1}{2}\left(\frac{GM_\bullet}{I}\right)^2 + 
\overline\Phi_p, \\
\overline\Phi_p &\equiv& \oint\frac{dw}{2\pi} \Phi = 
\frac{1}{2\pi}\int_0^{2\pi} dE\,(1-e\cos E)\,\Phi_p({\bf r}).
\end{eqnarray}
\end{subequations}
The final term replaces the mean anomaly $w$ by the eccentric anomaly $E$,
where $r=a(1-e\cos E)$ and the eccentricity is  $e=\sqrt{1-L^2/I^2}$.
After the averaging, $\overline{\cal H}$ is independent of $w$,
and $I$ is conserved, as is the semi-major axis $a$.
We are left with four variables and with $\overline\Phi_p$ as 
the effective Hamiltonian of the system. 

The spherically symmetric part of $\overline\Phi_p$ is
\begin{eqnarray}
\overline\Phi_s &=& F_\gamma(e)\,\Phi_s\left(\frac{a}{r_0}\right)^{2-\gamma}, \\ 
F_\gamma(e) &\equiv& {}_2F_1 \left( \left[-\frac{3-\gamma}{2},-\frac{2-\gamma}{2}\right], [1], e^2\right). \nonumber
\end{eqnarray}
A good approximation to $F_\gamma(e)$ is 
\begin{equation}  \label{alpha_appr}
F_\gamma(e) \approx 1+\alpha e^2\;, \quad 
\alpha=\frac{2^{3-\gamma}\Gamma(\frac{7}{2}-\gamma)}{\sqrt{\pi}\,\Gamma(4-\gamma)}-1
\end{equation}
which is exact for $\gamma=0$ and $\gamma=1$;
for $0\le\gamma<2$, $0<\alpha\le 3/2$.
When $\gamma>1$ and $e$ is close to 1, a better approximation is
\begin{equation}  \label{Equation:alphap_appr}
F_\gamma(e) \approx 1+\alpha + \alpha'(e^2-1) \;, \quad
\alpha' = \frac{2^{1-\gamma}(2-\gamma)}{\sqrt\pi} \frac{\Gamma(\frac{5}{2}-\gamma)}{\Gamma(3-\gamma)}.
\end{equation}
We adopt the latter expression in what follows.
Good approximations are
\begin{subequations}
\begin{eqnarray}
\alpha  &\approx& \textstyle \frac{3}{2}-\frac{79}{60}\gamma+\frac{7 }{20}\gamma^2-\frac{1}{30}\gamma^3, \\
\alpha' &\approx& \textstyle \frac{3}{2}-\frac{29}{20}\gamma+\frac{11}{20}\gamma^2-\frac{1}{10}\gamma^3.
\end{eqnarray}
\end{subequations}
Similar expressions can be found in \citet{Ivanov2005} and
\citet{PPS2007}.

Expressions for the orbit-averaged triaxial harmonic potential 
(excluding the spherical component) are derived in \citet{SambhusSridhar2000}.
Adopting their notation,
the orbit-averaged potential in our case becomes
\begin{subequations}
\begin{eqnarray}
\overline\Phi_p &=& \Phi_s\left(\frac{a}{r_0}\right)^{2-\gamma} (1+\alpha-\alpha'\ell^2) 
  + 2\pi G\rho_t T_x\,a^2 \times \\
  &\times& \left[ \frac{5}{2} - \frac{3}{2}\ell^2 + 
  \epsilon_b^{(t)} H_b(\ell,\ell_z,\varpi,\Omega) + \epsilon_c^{(t)} H_c(\ell,\ell_z,\varpi) \right] \nonumber 
\label{eq:PhiAveragea} \\
H_b &=& {\textstyle\frac{1}{2}} \left[ (5-4\ell^2)(c_{\varpi} s_\Omega + c_i c_\Omega s_{\varpi})^2 \right.\\
&+& \left. \ell^2(s_{\varpi} s_\Omega - c_i c_\Omega c_{\varpi})^2 \right], \nonumber \\
H_c &=& {\textstyle\frac{1}{4}} (1-c_i^2)[5-3\ell^2-5(1-\ell^2)c_{2\varpi}], \\
\epsilon_b^{(t)} & \equiv&  T_y/T_x-1 \;,\quad 
\epsilon_c^{(t)} \;\equiv\; T_z/T_x-1 .
\end{eqnarray}
\label{eq:PhiAverage}
\end{subequations}
The shorthand $s_x, c_x$ has been used for $\sin x, \cos x$.
We have defined the dimensionless variables
$\ell=L/I$ and $\ell_z=L_z/I$, both of which vary from 0 to 1; 
the orbital inclination $i$ is given by 
$\cos i \equiv \ell_z/\ell$ and the eccentricity by $e^2=1-\ell^2$.

\begin{figure*}[t]
\includegraphics{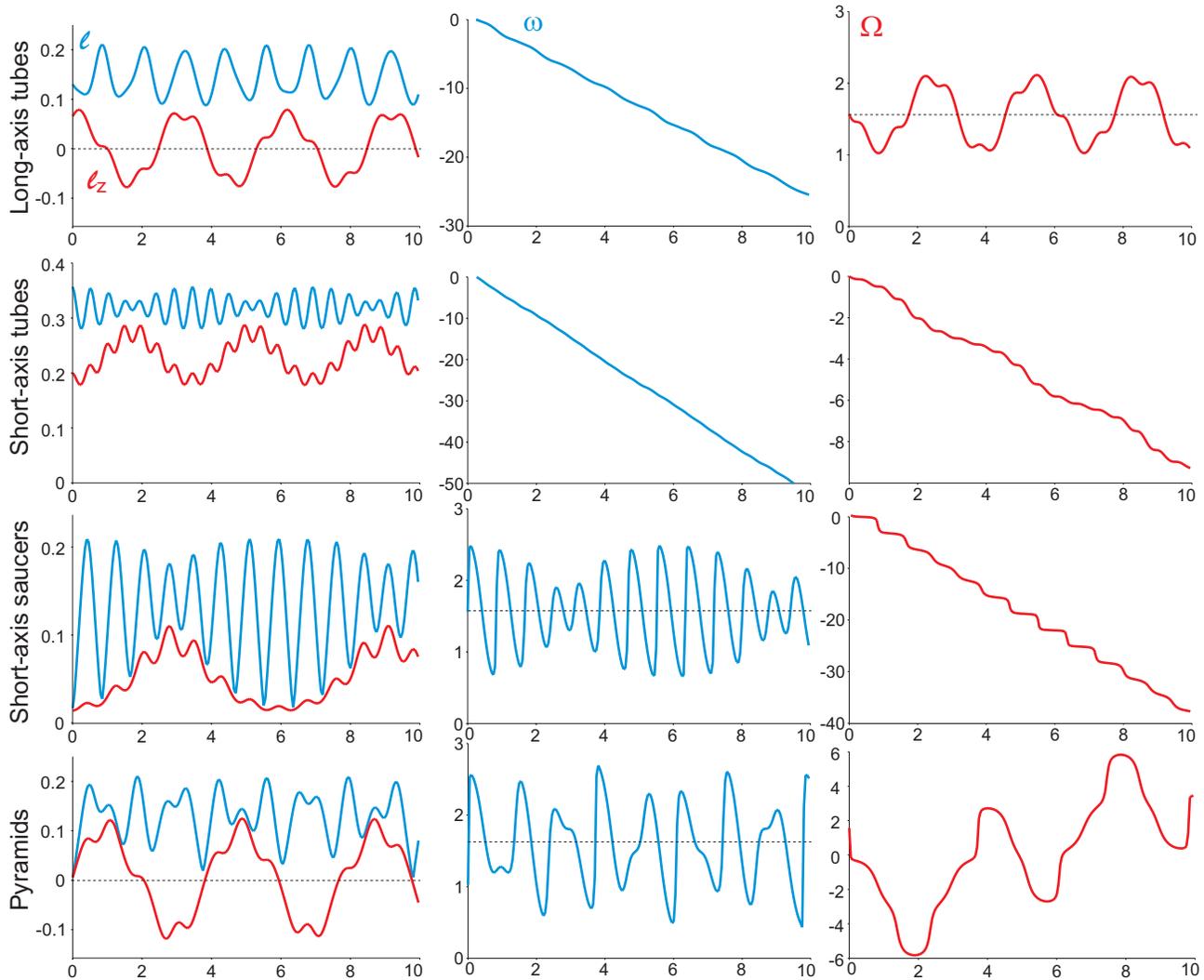}
\caption{Four classes of orbits around a BH in a triaxial nucleus.
{\it Left column}: time dependence of the dimensionless angular
momentum $\ell$ (top/blue) and its component $\ell_z$ along the short
axis of the figure (bottom/red). 
{\it Middle column}: argument of the periapse $\varpi$.
{\it Right column}: angle of nodes $\Omega$. 
} \label{fig:orbitclasses}
\end{figure*}

\begin{figure}[t]
\includegraphics{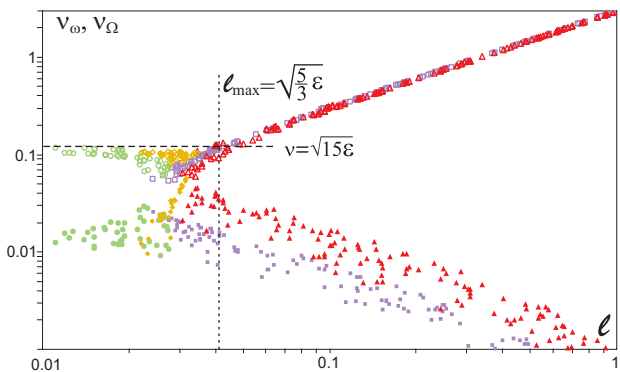}
\caption{Dependence of the characteristic frequencies $\nu_\varpi$ 
(in-plane precession) and $\nu_\Omega$ (nodal precession) 
on the value of the dimensionless angular momentum $\ell$.  
{\it Top (open) symbols}: $\nu_\varpi$; {\it bottom (filled) symbols}:
$\nu_\Omega$. 
{\it Magenta boxes}: LATs; 
{\it red triangles}: SATs;
{\it yellow diamonds}: saucers;
{\it green circles}: pyramids.
For large $\ell$ $\nu_\varpi \propto \ell$ 
and $\nu_\Omega \ll \nu_\varpi$, but for 
sufficiently low $\ell$ these two are comparable, 
which gives birth to the pyramid and saucer orbits. 
Vertical line denotes the threshold in $\ell$ (equation~\ref{ell_max}), 
horizontal line the characteristic frequency 
$\nu_{x0}$ (equation~\ref{nu0}).
Triaxiality coefficients were set to
$\epsilon_c=10^{-3}$, $\epsilon_b=0.4\epsilon_c$.
} \label{fig:frequencies}
\end{figure}

The first term in equation~(\ref{eq:PhiAveragea}), which arises
from the spherically-symmetric cusp, does not depend on the
angular variables, and has the same dependence on $\ell^2$ as the
corresponding term in the harmonic triaxial potential.
So we can sum up the coefficients at $\ell^2$ and renormalize
the triaxial coefficients $\epsilon_{b,c}$ to obtain the
same functional form of the Hamiltonian as in the purely
harmonic case.
Dropping an unnecessary constant term (depending only on $a$) and
defining a dimensionless time $\tau=\nu_p t$, where 
$\nu_p$ is characteristic rate of precession,
\begin{eqnarray}  \label{eq:eps_tilde}
\nu_p &\equiv& 2\pi G\rho_t T_x a^2 (1+A)/I \;, \\
A &\equiv& {\textstyle \frac{4\alpha'}{3(3-\gamma)(2-\gamma)T_x}} \frac{\rho_s}{\rho_t} \left(\frac{a}{r_0}\right)^{-\gamma},
\end{eqnarray}
we obtain the dimensionless Hamiltonian and the equations of motion
describing the perturbed motion:
\begin{subequations}
\label{eq:Averaged}
\begin{eqnarray}  
\label{eq:HAverage}
H &\equiv& \frac{\overline\Phi_p}{\nu_p I} = -\frac{3}{2}\ell^2 + \epsilon_b H_b + \epsilon_c H_c, \\
\label{eq:MotionAverage}
\frac{d\ell}{d\tau} &=& -\frac{\partial H}{\partial\varpi},\ 
\frac{d\varpi}{d\tau} = \frac{\partial H}{\partial\ell},\ 
\frac{d\ell_z}{d\tau} = -\frac{\partial H}{\partial\Omega},\ 
\frac{d\Omega}{d\tau} = \frac{\partial H}{\partial\ell_z}. \qquad\mathstrut
\end{eqnarray}
\end{subequations}
The renormalized triaxiality coefficients are
$\epsilon_{b,c} \equiv \epsilon_{b,c}^{(t)}/(1+A)$.

If there were no spherical component ($A=0$), this would reduce to the
purely harmonic triaxial case studied by \citet{SambhusSridhar2000}.\footnote{
Equation (12) of \cite{SambhusSridhar2000} for $\dot\ell$
lacks a minus sign in front of the first term.}
Adding the spherically symmetric cusp increases the rate of periapse
precession, while at the same time reducing the relative amplitude of the
triaxial terms; otherwise the form of the Hamiltonian is essentially
unchanged.

A more transparent expression for the precession frequency $\nu_p$ is
\begin{subequations}
\begin{eqnarray}  \label{eq:nu_p}
\nu_p &=& \nu_r \left(\frac{M_t(a)}{M_\bullet} \frac{3T_x}{2} + \frac{M_s(a)}{M_\bullet} \frac{2\alpha'}{3(2-\gamma)} \right) \\
M_t(a) &\equiv& \frac{4\pi}3 a^3\rho_t  \;\;,\quad  
M_s(a) \equiv \frac{4\pi}{3-\gamma} a^3 \rho_s \left(\frac{a}{r_0}\right)^{-\gamma}
\end{eqnarray}
\end{subequations}
where $M(a)$ denotes the mass
enclosed within radius $r=a$.
From equation~(\ref{eq:MotionAverage}), 
the precession rate of an orbit in 
the spherical cluster is 
$\nu_{\varpi} \equiv|d\varpi/dt|=3\ell\nu_p=3\sqrt{1-e^2}\,\nu_p$.
Near the BH\ influence radius, 
it is clear that $\nu_p \approx \nu_r$;
hence the orbit-averaged treatment,
which assumes only one ``fast'' variable,
is likely to break down at this radius.

We note that $\nu_{\varpi}\rightarrow 0$ as $e\rightarrow 1$.
For the very eccentric orbits that are the focus of this paper,
the rate of precession is much lower than for a typical, non-eccentric
orbit of the same energy.
This will turn out to be important, since the slow precession
allows torques from the triaxial part of the potential to build up.

\section{Orbital structure of the model potential}  \label{sec_analysis}
\subsection{General remarks}

The orbit-averaged Hamiltonian (\ref{eq:HAverage}) describes 
a dynamical system of two degrees of freedom. 
The trajectories must be obtained by numerical integration of the
equations of motion (\ref{eq:MotionAverage}).
We begin by making some qualitative points about the nature of the solutions.

In the absence of the triaxial terms in equation~(\ref{eq:HAverage}),
the effect of the distributed mass is to rotate the periapse angle 
$\varpi$ in a fixed plane;
this steadily rotating elliptic orbit fills an annulus.
The addition of a weak triaxial perturbation changes the rate
of in-plane precession slightly, and also causes the orbital plane 
itself to change, as described
by the last two terms in equation~(\ref{eq:MotionAverage}).

In general, two angular variables $\varpi$ and $\Omega$ can either
librate around fixed points or circulate, giving rise to four
basic families of orbits (Figure~\ref{fig:orbitclasses}).
Solutions to the equations of motion
that are characterized by circulation in both $\varpi$ and
$\Omega$ correspond to tube orbits about the short axis (SAT).
Motion that circulates in $\varpi$ but librates in $\Omega$ 
corresponds to tube orbits about the long axis (LAT).
Both types of orbit are qualitatively similar to the tube orbits 
that are generic to the triaxial geometry \citep{Schwarzschild1979}.
A subclass of the SAT orbits corresponds to motion
that circulates in $\Omega$ and librates in $\varpi$
\citep{SambhusSridhar2000,PoonMerritt2001}.
These orbits resemble cones, or saucers; similar orbits
exist also at $r\gg\rh$ in oblate or nearly oblate potentials
\citep{Richstone1982,LeesSchwarzschild1992}.

If the degree of triaxiality is small ($\epsilon_{b,c} \ll 1$), 
then as noted above, the dominant effect of the distributed 
mass is simply to induce a periapse shift,
at a rate $\nu_{\varpi} \equiv d\varpi/d\tau = -3\ell$.
If the additional mass is much less than the mass of the BH,
then on short time scales (comparable to the radial period $\nu_r^{-1}$) 
the orbit resembles a nearly closed ellipse.
On intermediate time scales (of order the precession time $\nu_{\varpi}^{-1}$), 
a steadily-rotating elliptic orbit fills an annulus in a fixed plane. 
On still longer time scales $\nu_\Omega^{-1}$, the orbital plane itself changes due to 
the torques from the triaxial potential. 
Similar considerations give rise to the concept of vector resonant relaxation 
\citep{RauchTremaine1996}.

The foregoing description is valid as long as the angular momentum is 
not too low.
Since the precession rate is proportional to $\ell$, 
for sufficiently low $\ell$ 
the intermediate and long time scales become comparable
(Figure~\ref{fig:frequencies}).
As a result, the triaxial torques can produce substantial
changes in $\ell$ (i.e. the eccentricity)
on a precession time scale  via the
first term in~(\ref{eq:MotionAverage}), and the circulation 
in $\varpi$ can change to libration. 
This is the origin of the pyramid orbits, which are unique to the
triaxial geometry \citep{MerrittValluri1999}.

\subsection{Pyramid orbits}  \label{sec_pyramids}

Of the four orbit families discussed above, the first three
were treated, in the orbit-averaged approximation, by 
\citet{SambhusSridhar2000}.
The fourth class of orbits, the pyramids, are three-dimensional 
analogs of the two-dimensional ``lens'' orbits discussed by 
\cite{SridharTouma1997}, also
in the context of the orbit-averaged equations.
An important property of the pyramid orbits
is that $\ell$ can come arbitrarily close to zero
\citep{PoonMerritt2001} and \citet{MerrittPoon2004}.
This makes the pyramids natural candidates for 
providing matter to BHs at the centers of galaxies.

Pyramid orbits can be treated analytically if the following two
additional aproximations are made: 
(1) the angular momentum is assumed to be small, $\ell^2\ll 1$;
(2) the triaxial component of the potential is assumed to be
small compared with the spherical component, i.e. $\epsilon_b, \epsilon_c\ll1$.
As shown below, these two conditions are consistent, in the sense 
that $\ell_\mathrm{max}^2\sim\epsilon_{b,c}$ for pyramid orbits.

Removing the second-order terms in 
$\epsilon_b,\epsilon_c$ and in $\ell^2$ 
from the orbit-averaged Hamiltonian (\ref{eq:HAverage}),
we find
\begin{equation}  \label{eq:HApprox}
H = -\frac{3}{2}\ell^2 + \frac{5}{2}\left[\epsilon_c(1-c_i^2)s_{\varpi}^2
+ \epsilon_b (c_{\varpi} s_\Omega + c_i s_{\varpi} c_\Omega)^2  \right]
\end{equation}
where again $c_i \equiv \cos i = \ell_z/\ell$.

Because an orbit described by (\ref{eq:HApprox}) is essentially
a precessing rod,
one expects the important variables to be the two that 
describe the orientation of the rod, and its eccentricity.
This argument led us to search for exact solutions to 
the equations of motion 
in terms of the Laplace-Runge-Lenz vector, or its 
dimensionless counterpart, the eccentricity vector, which point
in the direction of orbital periapse.
 
We therefore introduced new variables $e_x$, $e_y$ and $e_z$:
\begin{subequations}
\label{eq:exey}
\begin{eqnarray}  
e_x &=& \cos\varpi\,\cos\Omega - \sin\varpi\,\cos i\,\sin\Omega \\
e_y &=& \sin\varpi\,\cos i\,\cos\Omega + \cos\varpi\,\sin\Omega \\
e_z &=& \sin\varpi\,\sin i 
\end{eqnarray}
\end{subequations}
which correspond to components of a unit vector in the direction of 
the eccentricity vector. 
Of these, only two are independent, since $e_x^2+e_y^2+e_z^2=1$.
In terms of these variables, the Hamiltonian (\ref{eq:HApprox})
takes on a particularly simple form:
\begin{equation}  \label{eq:Hprime}
H = - \frac{3}{2}\ell^2 + \frac{5}{2}\left[ \epsilon_c - \epsilon_c e_x^2 - (\epsilon_c-\epsilon_b) e_y^2 \right].
\end{equation}
As expected, the Hamiltonian depends on only three variables:
$e_x$ and $e_y$, which describe the orientation of the orbit's
major axis, and the eccentricity $\ell$.

To find the equations of motion, we must switch to a Lagrangian formalism. 
Taking the first time derivatives of equations
(\ref{eq:exey}) and using 
equations~(\ref{eq:MotionAverage}), we find 
\begin{subequations}
\label{eq:exeydot}
\begin{eqnarray}
&&\dot e_x = 3\ell (\sin\varpi\,\cos\Omega + \cos\varpi\,\sin\Omega\,\cos i), \\
&&\dot e_y = 3\ell (\sin\varpi\,\sin\Omega - \cos\varpi\,\cos\Omega\,\cos i)
\end{eqnarray}
\end{subequations}
where $\dot e_x\equiv de_x/d\tau$ etc.
Taking second time derivatives, the variables describing 
the orientation and eccentricty of the orbit drop out, 
as desired, and the equations of motion for $e_x$ and $e_y$
can be expressed purely in terms of $e_x$ and $e_y$:
\begin{subequations}
\label{eq:exeyddot}
\begin{eqnarray}
\ddot e_x &=& -e_x\,6(H+3\ell^2)  \\
&=&  -e_x\,[ 30\epsilon_c - 6H - 30\epsilon_c e_x^2 - 30(\epsilon_c-\epsilon_b)e_y^2], \nonumber\\
\ddot e_y &=& -e_y\,6(H+3\ell^2-{\textstyle \frac{5}{2}}\epsilon_b) \\
&=&  -e_y\,[ 30\epsilon_c - 6H - 15\epsilon_b - 30\epsilon_c e_x^2 - 30(\epsilon_c-\epsilon_b)e_y^2]. \nonumber
\end{eqnarray}
\end{subequations}
From equations~(\ref{eq:exeydot}), 
$(\dot e_x,\dot e_y)=0$ implies $\ell=0$,
i.e. the eccentricity reaches one at the ``corners'' of the orbit.
These define the base of the pyramid.
Defining $(e_{x0},e_{y0})$ to be the values of $(e_x,e_y)$ when
this occurs,
vthe Hamiltonian has numerical value
\begin{equation}
H = \frac{5}{2}\epsilon_c - \frac{5}{2}\left[\epsilon_c e_{x0}^2 +
(\epsilon_c-\epsilon_b)e_{y0}^2\right].
\end{equation}

Equations~(\ref{eq:exeyddot}) have the form of coupled, nonlinear oscillators. 
Given solutions to these equations, the time dependence of the 
additional variables 
($\ell,\ell_z,\varpi,\Omega$) follows immediately from 
equations~(\ref{eq:exey}) and (\ref{eq:exeydot}):
\begin{subequations}
\begin{eqnarray*}
\ell^2 &=& \frac{\dot e_x^2 + \dot e_y^2 - (\dot e_x e_y - e_x \dot e_y)^2}{9 (1-e_x^2-e_y^2)} =
 \frac{1}{9} (\dot e_x^2 + \dot e_y^2 + \dot e_z^2),   \label{ell_exey} \\
\ell_z &=& (\dot e_x e_y - e_x \dot e_y)/3,  \label{ellz_exey} \\
\sin^2\varpi &=& \frac{1-e_x^2-e_y^2}{1-\ell_z^2/\ell^2} = \frac{e_z^2}{1-\ell_z^2/\ell^2}, \\
\dot e_z^2 &=& \frac{(e_x \dot e_x + e_y \dot e_y)^2}{1-e_x^2-e_y^2}\;,\quad
e_z^2 = 1-e_x^2-e_y^2,
\end{eqnarray*}
\end{subequations}
and a quite lengthy expression for $\Omega$ which we choose not to reproduce here.

\begin{figure*}[t]
\includegraphics[angle=-90,width=0.90\textwidth]{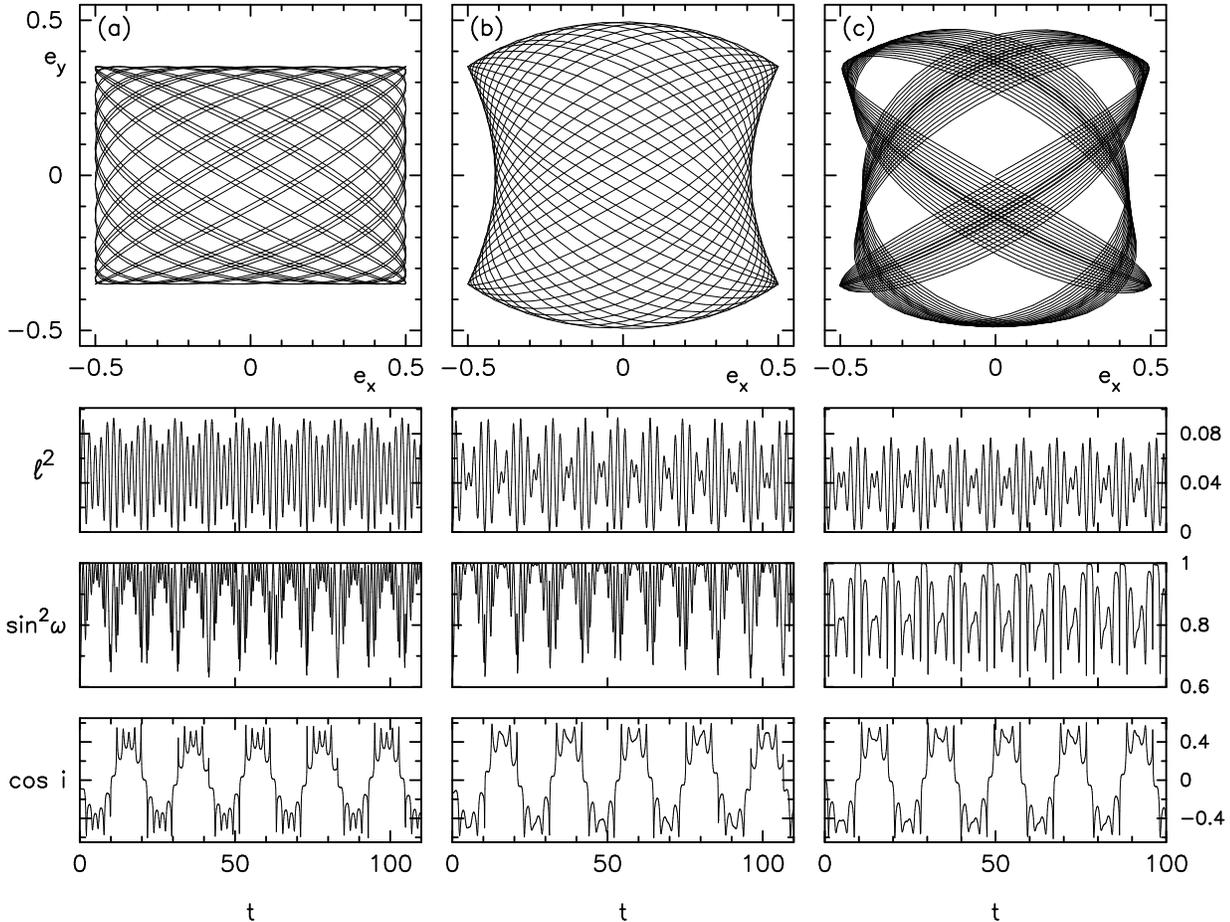}
\caption{A pyramid orbit, in three approximations.
Each orbit has the same $(e_{x0}, e_{y0}) = (0.5,0.35)$.
(a) The simple harmonic oscillator (SHO) approximation, 
equations~(\ref{eq:Highe}),
valid for small $\ell^2$, $(\epsilon_b, \epsilon_c)$ and $(e_{x0}, e_{y0})$.
(b) From equation~(\ref{eq:exeyddot}), 
which does not assume small $(e_{x0}, e_{y0})$.
(c) From the full orbit-averaged equations~(\ref{eq:Averaged}), which
does not assume small $\ell^2$, $\epsilon$ or ($e_{x0}, e_{y0}$). 
Aside from the fact that the latter orbit is fairly close to a
$5:2$ resonance, the correspondence between the physically important properties
of the approximate orbits is good. 
The triaxiality parameters are $(\epsilon_b, \epsilon_c)= (0.0578, 0.168)$,
corresponding to a pyramid orbit with $a=0.1r_0$
in a nucleus with triaxial axis ratios $(0.5,0.75)$, density ratio
$\rho_t(r_0)/\rho_s(r_0) = 0.1$, and $\gamma=1$.
The frequencies for the SHO case are $\nu_{x0}=1.59$, $\nu_{y0}=1.28$
(equation~\ref{nu0}); 
frequencies for planar orbits with the same $e_x$ and $e_y$ amplitudes 
are 1.48 and 1.24 respectively (equation~\ref{eq:periods}).
} \label{fig:orbits}
\end{figure*}

In the limit of small amplitudes, $(e_x,e_y) \ll 1$
(which corresponds to $H\approx \frac{5}{2}\epsilon_c$),
the oscillations are harmonic and uncoupled, 
with dimensionless frequencies 
\begin{equation}  \label{nu0}
\nu_{x0} \equiv \sqrt{15\epsilon_c} \;,\;\; 
\nu_{y0} \equiv \sqrt{15(\epsilon_c-\epsilon_b)}.
\end{equation}
The apoapse traces out a 2d Lissajous figure
in the plane perpendicular to the short axis of the triaxial figure.
This is the base of the pyramid (e.g. \citet{MerrittValluri1999}, Figure~11).
The solutions in this limiting case are
\begin{subequations}
\label{eq:Highe}
\begin{eqnarray}
e_x(\tau) &=& e_{x0}\cos(\nu_{x0}\tau + \phi_x), \\ 
e_y(\tau) &=& e_{y0}\cos(\nu_{y0}\tau + \phi_y), \\
\ell^2(\tau) &=& \ell_{x0}^2\sin^2(\nu_{x0}\tau+\phi_x) + 
\ell_{y0}^2\sin^2(\nu_{y0}\tau+\phi_y)  \nonumber
\end{eqnarray}
\end{subequations}
where $\phi_x,\phi_y$ are arbitraty constants and 
\begin{equation}
\ell_{x0}=\nu_{x0} e_{x0}/3,\ \ \ \ \ell_{y0}=\nu_{y0} e_{y0}/3.
\label{eq:ell0}
\end{equation}
Figure~\ref{fig:orbits}a plots an example.

Equations~(\ref{eq:Highe}) describe integrable motion.
Remarkably, it turns out that the more general (anharmonic, coupled)
equations of motion~(\ref{eq:exeyddot}) are  integrable as well.
The first integral is $H$;
an equivalent, but nonnegative, integral is $U$ where
\begin{equation}
U \equiv 15\epsilon_c-6H = \nu_{x0}^2 e_x^2 + \nu_{y0}^2 e_y^2 + 
 (\dot e_x^2 + \dot e_y^2 + \dot e_z^2).
\label{eq:U}
\end{equation}
The second integral is obtained after multiplying the 
first of equations~(\ref{eq:exeyddot}) by $15\epsilon_c \dot e_x$, 
the second by $15(\epsilon_c-\epsilon_b) \dot e_y$, 
and adding them to obtain a complete differential. 
The integral $W$ is then
\begin{subequations}
\begin{eqnarray}
W &=& \nu_{x0}^2 (\dot e_x^2 + \nu_x^2 e_x^2 - \nu_{x0}^2 e_x^4) + 
      \nu_{y0}^2 (\dot e_y^2 + \nu_y^2 e_y^2 - \nu_{y0}^2 e_y^4)   \nonumber \\
  &-& 2\nu_{x0}^2\nu_{y0}^2\, e_x^2 e_y^2, \\
\nu_x^2 &\equiv& U + \nu_{x0}^2 \;,\quad
\nu_y^2  \equiv  U + \nu_{y0}^2.
\label{eq:W}
\end{eqnarray}
\end{subequations}
The existence of two integrals ($U,W$), 
for a system with two degrees of freedom,
demonstrates regularity of the motion.

Regular motion can always be expressed in terms of action-angle variables.
The period of the motion, in each degree of freedom, is then given
simply by the time for the corresponding angle variable to increase
by $2\pi$.
We were unable to derive analytic expressions for the action-angle
variables corresponding to the two-dimensional 
motion described by equations~(\ref{eq:exeyddot}).
However the periods of oscillation of the {\it planar} orbits 
($e_x=0$ or $e_y=0$) described
by these equations are easily shown to be
\begin{eqnarray}\label{eq:periods}
\nu_{x0}P(e_{x0}) &=& 4K\left(e_{x0}^2\right)\ \  (e_y=0),\\ 
\nu_{y0}P(e_{y0}) &=& 4K\left(e_{y0}^2\right)\ \  (e_x=0) \nonumber
\end{eqnarray}
where $K(\alpha)$ is the complete elliptic integral:
\begin{equation}
K(\alpha) = \int_0^{\pi/2} \left(1-\alpha\sin^2x\right)^{-1/2}\;dx. \nonumber
\end{equation}
For small $\alpha$, $K\approx \pi/2$ and $P\approx 2\pi/\nu_0$.
As $\alpha\rightarrow 1$, $K\rightarrow\infty$; this corresponds
to a pyramid that precesses from the $z$ axis all the way to  the
($x,y$) plane.
The oscillator is ``soft'': increasing the amplitude  increases 
also the period.
Figure~\ref{fig:orbits} shows comparison of orbits with the same 
initial conditions, calculated in three different approximations.

Pyramid orbits can be seen as analogs of regular box orbits in 
triaxial potentials \citep{Schwarzschild1979},
with three independent oscillations in each coordinate.
Like box orbits, they do not conserve the magnitude of the
sign of the angular momentum about any axis.
The difference is that a BH in the center serves as a kind of 
``reflecting boundary'', so that a pyramid orbit is reflected by $180^\circ$ 
near periapsis, instead of continuing its way to the other
side of $x-y$ plane as a box orbit would do.

\subsection{The complete phase space of eccentric orbits}  \label{sec_phaseecc}

\begin{figure}[t] 
\includegraphics{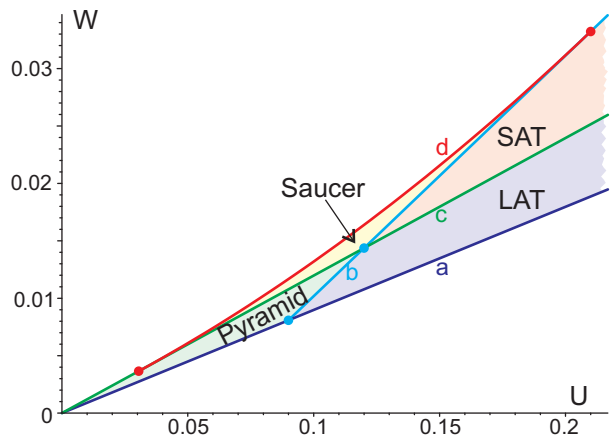}
\caption{Regions in the $U-W$ plane occupied by the different orbit families. 
{\it Lower (dark blue) line} ($a$), equation~(\ref{UW_low});
{\it upper (red) curve} ($d$), equation~ (\ref{UW_uppHC});
{\it green line} ($c$), equation~(\ref{UW_uppLAT}); 
{\it light blue line} ($b$), equation~(\ref{UW_pyrLAT}); 
{\it red points}, equation~(\ref{UW_HCpoints});
{\it blue points}, equation~(\ref{UW_PyrPoints}). 
Plotted for $\epsilon_b=0.002, \epsilon_c=0.008$. 
} \label{fig:UW}
\end{figure}

While our focus is on the pyramid orbits,
the low-angular-momentum Hamiltonian (\ref{eq:HApprox}) 
also supports orbits from other families.
In this section, we complete the discussion of the phase space 
described by equation~(\ref{eq:HApprox}), by delineating the regions in the 
$U-W$ plane that are occupied by each of the four orbit families 
(Figure~\ref{fig:UW}).

Pyramids and LATs both resemble distorted rectangles in the $e_x, e_y$ plane. 
The corner points of this region correspond to $\dot e_x=\dot e_y=0$.
Evaluating the two integrals at a corner (denoted by the subscript $0$) 
gives
\begin{subequations}  \label{UWlat}
\begin{eqnarray}
U &=& \nu_{x0}^2\, e_{x0}^2 + \nu_{y0}^2\, e_{y0}^2 + \dot e_{z,0}^2 \,, \\
W &=& \nu_{x0}^4\, e_{x0}^2 + \nu_{y0}^4\, e_{y0}^2 + (U-\dot e_{z,0}^2) \dot e_{z,0}^2 \,. 
\end{eqnarray}
\end{subequations}
The difference between pyramids and LATs arises from the last term: 
corner points of pyramid orbits 
correspond to $\ell^2=0$ and hence (from (\ref{ell_exey})) 
to $\dot e_{z,0}=0$. 
For LATs the condition is, conversely, $\dot e_{z,0}^2>0$, and $e_{z,0}=0$ 
(hence $e_{y0}^2 = 1-e_{x0}^2$). 
Analyzing these expressions, we find that for pyramids and LATs 
the lower and upper boundaries for $W$ given $U$ are
\begin{eqnarray}
W &=& \nu_{y0}^2\, U\,,   \label{UW_low} \\
W &=& \nu_{x0}^2\, U\,,   \label{UW_uppLAT} 
\end{eqnarray}
and the boundary between pyramids and LATs is given by 
\begin{equation}
W = (\nu_{x0}^2+\nu_{y0}^2)\, U - \nu_{x0}^2\,\nu_{y0}^2 .  
\label{UW_pyrLAT}
\end{equation}
Pyramids lie above and to the left of this line in the $U-W$ plane, 
while LATs are below and to the right. 
The intersection of this line with (\ref{UW_low}) and 
(\ref{UW_uppLAT}) occurs at the points
\begin{subequations}  \label{UW_PyrPoints}
\begin{eqnarray}  
U&=&\nu_{y0}^2,\; W=\nu_{y0}^4\,,\\ 
U&=&\nu_{x0}^2,\; W=\nu_{x0}^4\,.
\end{eqnarray}
\end{subequations}
These points constitute the leftmost bound for LATs and the 
rightmost bound for pyramids respectively.

Short-axis tubes and saucers resemble distorted rectangular regions 
in the $e_x,e_z$ plane.
Again, the corner points (with subscript 0) are defined to have 
$\dot e_x=\dot e_z=0$ and $e_y=0$, 
with $\dot e_y^2>0$, and therefore 
\begin{subequations}  \label{UWsat}
\begin{eqnarray}  
U &=& \nu_{x0}^2\, e_{x0}^2 + \dot e_{y0}^2  \,,\\
W &=& \nu_{x0}^2\, U + (\nu_{x0}^2 e_{x0}^2 + \nu_{y0}^2-\nu_{x0}^2)\, \dot e_{y0}^2 \,.
\end{eqnarray}
\end{subequations}
Both these families have $W\ge \epsilon_c U$, 
i.e. lie above the line (\ref{UW_uppLAT}).
SAT orbits intersect the plane $e_z=0$, 
so we can set $e_{x0}=1$ in (\ref{UWsat}). 
(Alternatively, for SATs, both angles circulate, 
so we can set $\varpi=\Omega=0$, which again gives $e_x=1$).
We then find that SATs lie below the line (\ref{UW_pyrLAT}).

On the other hand, saucers never reach $e_z=0$ 
(since for them $\sin^2 \varpi > 0$), 
so that they lie above the line (\ref{UW_pyrLAT}). 
To obtain the upper limit for $W$ at fixed $U$, we substitute 
$\dot e_{y0}^2$ from the first equation in (\ref{UWsat}) in the second, 
and then seek a maximum of $W$ 
with respect to $e_{x0}$ at fixed $U$. This gives 
\begin{equation}
W = \nu_{x0}^2\, U + (U+\nu_{y0}^2-\nu_{x0}^2)^2/4   \label{UW_uppHC}.
\end{equation}
This curve intersects  (\ref{UW_uppLAT}) and (\ref{UW_pyrLAT}) in the points
\begin{subequations}  \label{UW_HCpoints}
\begin{eqnarray}
U=\nu_{x0}^2-\nu_{y0}^2,\; W=\nu_{x0}^2 U \,,\\ 
U=\nu_{x0}^2+\nu_{y0}^2,\; W=\nu_{x0}^2 U + \nu_{y0}^4 \,,
\end{eqnarray}
\end{subequations}
which define the left- and rightmost bounds for the saucer region.

All these criteria are summarized in Figure~\ref{fig:UW}. 
In particular, pyramid orbits exist in the following cases:
\begin{itemize}
\item for $0\le H \le \frac{5}{2}\epsilon_b$ they coexist with LATs;
\item for $0\le H \le \frac{5}{2}(\epsilon_c-\epsilon_b)$ they coexist with SAT saucers;
\item above these values they are the only population for $H\le \frac{5}{2}\epsilon_c$, which is the maximum allowed value of $H$.
\item below $H<0$ pyramids do not exist 
(this is easily seen from equation~(\ref{eq:HApprox}): 
since the term in square brackets is always non-negative, it is impossible to 
have $\ell^2=0$ when $H<0$).
\end{itemize}

\begin{figure}[t] 
\includegraphics{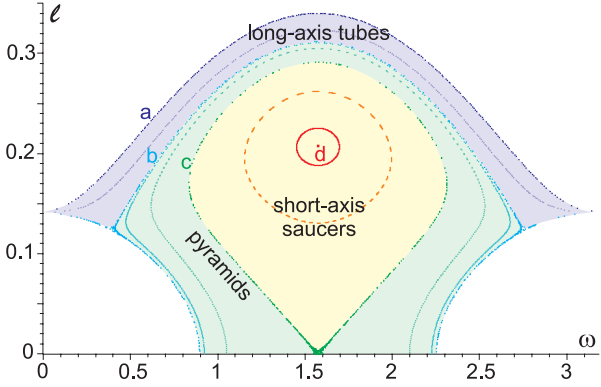}
\caption{Poincar\'e section for $\ell, \varpi$ plotted at $\Omega=\pi/2$ for 
energy $H=0.02$
($\epsilon_b=0.99^{-2}-1, \epsilon_c=0.96^{-2}-1$)
showing the three possible types of orbit:
LATs, pyramids and SATs/saucers. 
This figure adopts the same potential parameters as
Figure~5b in \cite{SambhusSridhar2000}, but those authors
chose $H=-0.02$ which precludes pyramid orbits.
Boundaries are marked by the same letters as in Figure~\ref{fig:UW}.
} \label{fig:PS}
\end{figure}

Figure~\ref{fig:PS} shows Poincar\'e surfaces of section for $\Omega=\pi/2$ 
and $0<H<\frac{5}{2}\epsilon_b$.
The three families of orbits are delineated.

Since $H$ is an integral of the motion, the maximum allowed value of 
$\ell^2$ can not exceed
\begin{equation}  \label{ell_max}
\ell_\mathrm{max}^2(H)=\frac{5\epsilon_c-2H}{3+4\epsilon_c-\epsilon_b} \approx \frac{1}{3}(5\epsilon_c-2H).
\end{equation}
The latter approximate expression is immediately seen from the
simplified Hamiltonian (\ref{eq:HApprox}), while the former comes from
the exact Hamiltonian (\ref{eq:Averaged}).
However, it does not follow that an orbit with a sufficiently low 
{\sl instantaneous} value of the angular momentum
is necessarily a pyramid: both tube families can also have arbitrarily low 
$\ell$. 
The principal distinction is that {\sl any} pyramid orbit can achieve
arbitrarily low $\ell$ (that is, the lower bound is $\ell=0$), 
while tube orbits always have $0<\ell_\mathrm{min}^2 \le \ell^2$ (however small 
$\ell_\mathrm{min}$ may be, it is strictly positive).

We now return from the simplified Hamiltonian (\ref{eq:HApprox}) to 
the full Hamiltonian (\ref{eq:Averaged}), i.e. we no longer 
require $\epsilon$ to be small. 
The full Hamiltonian retains all the qualitative properties of the 
simplified system but requires numerical integration of the equations 
of motion (\ref{eq:Averaged}) to determine orbit classes.

To quantify the overall fraction of pyramid orbits in a given potential, 
one should uniformly sample the phase space for all
four variables and determine the orbit class for each initial condition. 
From equation~(\ref{ell_max}), we can restrict ourselves to values of 
$\ell^2 \le \ell_\mathrm{max}^2(0) = \frac{5\epsilon_c}{3+4\epsilon_c-\epsilon_b}$ 
(but we must take care not to filter out initial conditions corresponding to 
$H<0$).

We calculated the proportions of the $\ell_\mathrm{max}^2$-restricted fraction of 
phase space occupied by each family of orbits. 
Initial conditions were drawn randomly for $10^4$ points 
(with uniform distribution in $\ell^2 \in [0..\ell^2_{max}]$, in $\ell_z\in[0..\ell]$, and in $\varpi,\Omega \in [0,\frac{\pi}2]$).
The proportions were found to depend very weakly on $\epsilon_c$ 
if $\epsilon_c \ll 1$. 
To elucidate the dependence on  $\epsilon_b/\epsilon_c$ 
(the degree of triaxiality) we took 15 values in the range ($0.001-0.999$).

\begin{figure*}[t] 
\includegraphics{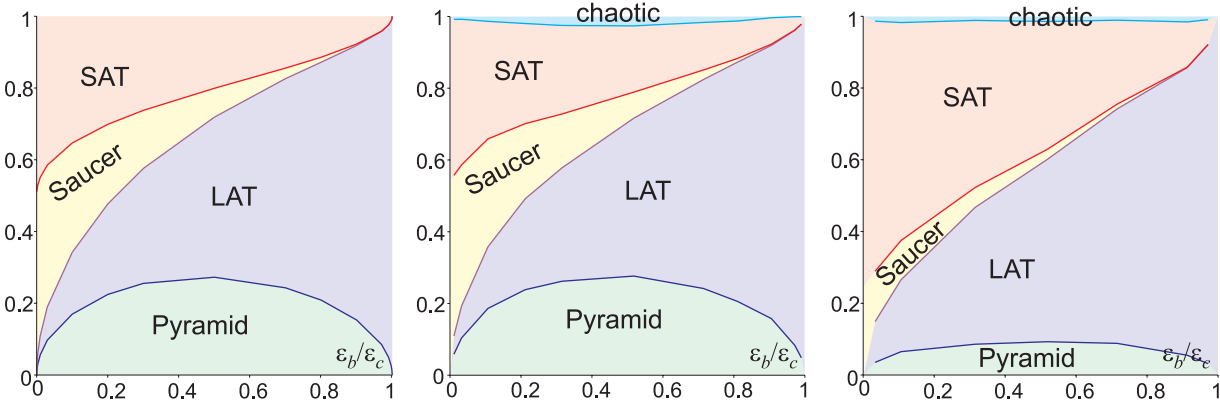}
\caption{Proportions of the restricted part of phase space 
(defined by $\ell^2<\ell_\mathrm{max}^2(0)$) that are occupied by the 
major orbit families:
LATs, pyramids, SATs, and SAT/saucers, as a function of the
ratio $\epsilon_b/\epsilon_c$. 
{\it Left}: analytic estimates from the simplified orbit-averaged Hamiltonian 
(\ref{eq:HApprox}) for $\epsilon_c\to 0$;
{\it middle}: orbit-averaged Hamiltonian (\ref{eq:Averaged}) 
for $\epsilon_c=0.1$;
{\it right}: real-space integration for orbits with semimajor axis 
$a=\rh$ (equal to the BH influence radius) and 
$\epsilon_c=0.1$.   
} \label{fig:proportion}
\end{figure*}
We found that the relative fraction $\eta$ of pyramids among {\sl low-}$\ell$ 
orbits is almost independent of $\epsilon_c$ (Figure~\ref{fig:proportion}):
\begin{equation}  \label{eta_pyr}
\eta \approx 0.28 \sqrt{4\frac{\epsilon_b}{\epsilon_c}
\left(1-\frac{\epsilon_b}{\epsilon_c}\right)}.
\end{equation} 
The fraction of pyramids among {\sl all} orbits is $\tilde\eta = \eta\ell_\mathrm{max}^2(0) = \frac{5}{3}\epsilon_c\,\eta$.
For comparison, the left panel of Figure~\ref{fig:proportion}
shows the results obtained using the simplified Hamiltonian (\ref{eq:HApprox}) 
and the analytical classification scheme 
described above, while the middle panel, made for $\epsilon_c=0.1$, 
shows almost the same behavior, with the addition of 
a small number of chaotic orbits.
We note that for $\epsilon_c \sim 1$ the phase space becomes largely chaotic.

These estimates of the relative fraction of pyramid orbits are 
directly applicable to a galaxy with an isotropic distribution 
of stars at any energy. 
This assumption may not be valid, for example, in the case of induced 
tangential anisotropy following the merger of supermassive BHs 
\citep{LivingReviews}.

One can ask a different question: 
if we know the instantaneous value of an orbit's eccentricity 
and orientation, what can we conclude about the orbit class? 
It is clear that without knowledge of the derivatives of $e_{x,y}$ 
the answer will only  be probabilistic.
It turns out that the probability $p$ for an orbit with 
``sufficiently high'' eccentricity (i.e. with 
$\ell^2 \le \ell_\mathrm{max}^2$) to be a pyramid depends mostly on the
$z$ component of the eccentricity vector:
$p \approx 0.7 \sqrt{4\frac{\epsilon_b}{\epsilon_c}(1-\frac{\epsilon_b}{\epsilon_c})}\, e_z^{1.5}$
(here the normalization comes from the total number of pyramids among {\sl low-}$\ell$ orbits). 
That is, an orbit lying in the plane defined by the long and intermediate axes of the potential is certainly not a pyramid, 
and the highest probabilitiy occurs for orbits directed toward the short axis.

\subsection{Large $\ell$ limit}  \label{sec_largel}

In the previous sections we considered the case $\epsilon_{b,c} \ll 1$ and 
$\ell^2 \sim \epsilon_c$, which allowed a simplification leading to integrable equations.

In the opposite case, when $\epsilon_{b,c} \ll \ell^2 \lesssim 1$, 
the frequency of in-plane precession, $\nu_\varpi$, is much greater than 
the rates change of $\Omega$ and $i$ (Figure~\ref{fig:frequencies}).
In this limit we can carry out a second averaging of the 
Hamiltonian~(\ref{eq:HAverage}), this time over $\varpi$.
Thus
\begin{eqnarray}\label{eq:Hav}
\langle H\rangle &=& \frac{1}{2\pi} \int_0^{2\pi} H d\varpi 
= -\frac{3}{2}\ell^2 \\
&+& \frac{5-3\ell^2}{4} \left[\epsilon_b\left(s_\Omega^2 + c_i^2 c_\Omega^2\right) 
 +  \epsilon_c\left(5-3\ell^2\right)\left(1- c_i^2\right) \right] . \nonumber
\end{eqnarray}
On timescales $T \gtrsim \nu_p^{-1}$ the orbit resembles an 
annulus that lies in the plane 
defined by the angles $i$ and $\Omega$. 
The only remaining equations of motion are those that describe
the change in orientation of the orbital plane:
\begin{eqnarray}
\frac{d\ell_z}{d\tau} &=& -\frac{\epsilon_b}{4}\left(5-3\ell^2\right)\left(1-c_i^2\right)s_{2\Omega}, \\
\ell\frac{d\Omega}{d\tau} &=& 
\frac{\epsilon_b}{2}\left(5-3\ell^2\right) c_\Omega^2c_i - 
\frac{\epsilon_c}{2}\left(5-3\ell^2\right) c_i.
\nonumber
\end{eqnarray}
One expects the natural variables in this case to be the components
of the angular momentum:
\begin{eqnarray}
\ell_x &=& \ell\sin i\sin\Omega,\\
\ell_y &=& \ell\sin i\cos\Omega,\nonumber \\
\ell_z &=& \ell\cos i \nonumber
\end{eqnarray}
and $\ell_x^2+\ell_y^2+\ell_z^2=\ell^2 = $ constant.
In terms of these variables, the Hamiltonian is
\begin{equation}\label{eq:Hel}
\langle H\rangle = -\frac{3}{2}\ell^2 +
\frac{(5-3\ell^2)}{4\ell^2} \left[ \epsilon_b \left(\ell^2-\ell_y^2\right) 
 + \epsilon_c\left(\ell^2-\ell_z^2\right) \right].
\end{equation}
After some algebra, one finds the equations of motion:
\begin{eqnarray}
\dot\ell_x &=& -\frac{1}{2}\left(\epsilon_c-\epsilon_b\right)
\left(5-3\ell^2\right)\frac{\ell_y\ell_z}{\ell^2}, \\
\dot\ell_y &=& \frac{\epsilon_c}{2}\left(5-3\ell^2\right)
\frac{\ell_x\ell_z}{\ell^2}, \nonumber \\
\dot\ell_z &=& -\frac{\epsilon_b}{2}\left(5-3\ell^2\right)
\frac{\ell_x\ell_y}{\ell^2} \nonumber
\end{eqnarray}
(only two of which are independent).
These can be written
\begin{eqnarray}
\frac{\mathrm{d}\vec\ell}{\mathrm{d}\tau} &=& \mathbf{T}\times\vec\ell,
\\
\mathbf{T} &=& \frac{5-3\ell^2}{2\ell^2}\left(
\begin{array}{c}
0\\
\epsilon_b\ell_y\\
\epsilon_c\ell_z
\end{array}
\right).\nonumber
\end{eqnarray}

Conservation of the Hamiltonian (\ref{eq:Hel}) implies
$$
\epsilon_b\ell_y^2 + \epsilon_c\ell_z^2 = \mathrm{constant} = C.
$$
This is an elliptic cylinder; the axis is parallel to the $\ell_x$-axis,
and the ellipse is elongated in the direction of the $\ell_y$-axis.
In addition, we know that
$$
\ell_x^2 + \ell_y^2 + \ell_z^2 = \mathrm{constant} = \ell^2
$$
which is a sphere.
So, the motion lies on the intersection of a sphere with an elliptic
cylinder.
There are two possibilities.

\noindent 1. $\ell^2 > C/\epsilon_b$. 
In this case, the cylinder intersects the sphere in a deformed
ring that circles the $\ell_x$-axis. This corresponds to a LAT orbit.

\noindent 2. $\ell^2 < C/\epsilon_b$.
In this case, the locus of intersection is a deformed ring
about the $\ell_z$-axis. This orbit is a SAT.

\noindent
In other words, precession of the angular momentum vector
can be either about the short or
long (not intermediate) axes of the triaxial ellipsoid.

\section{Capture of pyramid orbits by the BH}  \label{sec_capture}

As we have seen, pyramid orbits can attain arbitrarily low 
values of the dimensionless angular momentum $\ell$.
The BH tidally disrupts or captures stars with angular momentum less than a certain critical value $L_\bullet$, 
or -- in dimensionless variables -- $\ell_\bullet \equiv L_\bullet/I(a)$.
We can express $\ell_\bullet$ in terms of the capture radius $r_t$,
the radius at which a star is either tidally disrupted or swallowed. 
For BH masses greater than $\sim 10^8~M_\odot$, main sequence stars 
avoid disruption and  $r_t \approx r_\mathrm{Schw}\equiv 2GM_\bullet/c^2$;
for smaller $M_\bullet$, tidal disruption occurs outside the
Schwarzschild radius;
e.g. at the center of the Milky Way, $r_t\approx 10 r_\mathrm{Schw}$
for solar-type stars.
Defining $r_t=\Theta r_\mathrm{Schw}$
and writing $L_\bullet^2 \approx G\mh r_t$,
then gives
\begin{equation}  \label{ell_bullet}
\ell_\bullet^2 = \Theta\, \frac{r_\mathrm{Schw}}{a}  \approx
 10^{-5}\Theta \left(\frac{M_\bullet}{10^8M_\odot}\right) \left(\frac{a}{1\ \mathrm{pc}}\right)^{-1}.
\end{equation}

We note the following property of the pyramid orbits: 
as long as the frequencies of $e_x$ and $e_y$ oscillation are 
incommensurate, the vector $(e_x,e_y)$ fills densely the whole available area, which has the form of distorted rectangle.
The corner points correspond to zero angular momentum, and the ``drainage area'' is similar to four holes 
in the corners of a billiard table. 

Unless otherwise noted, in this section we adopt the simple harmonic 
oscillator (SHO) approximation to the ($e_x,e_y$) motion,
that is, we use the simplified Hamiltonian (\ref{eq:HApprox}) 
and its solutions (\ref{eq:Highe});
these orbits have $e_x^2+e_y^2 \ll 1$ and they  
form a rectangle in the $e_x-e_y$ plane, with sides $2e_x, 2e_y$. 
As long as the motion is integrable, the results for arbitrary pyramids 
with $e_x, e_y \lesssim 1$ will be qualitatively similar.
Quantitative results may be obtained by numerical analysis and are 
presented near the end of this section.

\begin{figure}[t]
\includegraphics{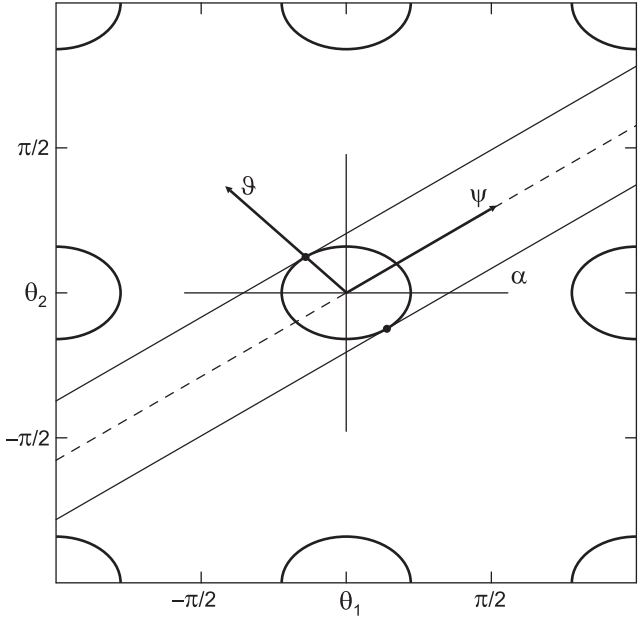}
\caption{Two-torus describing oscillations of ($e_x,e_y$) for a pyramid orbit.
The ellipses correspond to regions near the four corners of the
pyramid's base where $\ell\le\ell_\bullet$.
In the orbit-averaged approximation, 
trajectories proceed smoothly along lines parallel to the solid lines,
with slope $\tan\alpha = \nu_y/\nu_x$.
In reality, successive periapse passages occur at discrete intervals, 
once per radial period.
} \label{fig:phase}
\end{figure}

Figure~\ref{fig:phase} shows a two-torus describing oscillations in
($e_x,e_y$) for a pyramid orbit.
In the SHO approximation, solutions are given by (\ref{eq:Highe}).
If the two frequencies $\nu_{x0}, \nu_{y0}$ are incommensurate, the motion will fill the torus.
In this case, we are free to shift the time coordinate
so as to make both phase angles $(\phi_1, \phi_2)$ zero, yielding
\begin{subequations}
\begin{eqnarray}
\ell^2(\tau) &=& \ell_{x0}^2\sin^2(\nu_{x0}\tau) + 
\ell_{y0}^2\sin^2(\nu_{y0}\tau) \\
&=& \ell_{x0}^2\sin^2\theta_1 + \ell_{y0}^2\sin^2\theta_2
\end{eqnarray}
\label{eq:centered}
\end{subequations}
where $\theta_1=\nu_{x0}\tau, \theta_2=\nu_{y0}\tau$.
(In the case of exact commensurability, i.e.
$m_1\nu_{x0}+m_2\nu_{y0}=0$ with ($m_1,m_2$) integers,
the trajectory will avoid certain
regions of the torus and such a shift may not be possible.)
In the SHO approximation, $\nu_{x0}=\sqrt{15\epsilon_c},
\nu_{y0}=\sqrt{15(\epsilon_c-\epsilon_b)}$  (\ref{nu0}).
More generally, integrable motion will
still be representable as uniform motion on the torus
but the frequencies and the relations
between $\ell$ and the angles will be different.

Stars are lost when $\ell(\theta_1,\theta_2)\le \ell_{\bullet}$.
Consider the loss region centered at $(\theta_1,\theta_2)=(0,0)$.
This is one of four such regions, of equal size and shape, that correspond
to the four corners of the base of the pyramid.
For small $\ell_\bullet$, 
the loss region is approximately an ellipse,
\begin{equation}
\frac{\ell_{x0}^2}{\ell_\bullet^2}\theta_1^2 + 
\frac{\ell_{y0}^2}{\ell_\bullet^2}\theta_2^2 \lesssim 1. 
\end{equation}
The area enclosed by this ``loss ellipse'' is
\begin{equation}
\pi \frac{\ell_\bullet^2}{\ell_{x0}\ell_{y0}}.
\end{equation}
There are four such regions on the torus; together, they
constitute a fraction
\begin{equation}
\mu=\frac{1}{\pi} \frac{\ell_\bullet^2}{\ell_{x0}\ell_{y0}}
\label{eq:fraction}
\end{equation}
of the torus.

Stars move in the $(\theta_1,\theta_2)$ plane along
lines with slope $\tan\alpha=\nu_{y0}/\nu_{x0}$,
at an angular rate of $\sqrt{\nu_{x0}^2+\nu_{y0}^2}$.
Since periapse passages occur only once per radial
period, a star will move a finite step in the
phase plane between encounters with the BH.
The dimensionless time between successive periapse passages is
$\Delta t = 2\pi\nu_p/\nu_r$.
The angle traversed during this time is
\begin{equation}
\Delta\theta = 2\pi(\nu_p/\nu_r)\sqrt{\nu_{x0}^2 + \nu_{y0}^2} .
\end{equation}
The rate at which stars move into one the four loss ellipses is
given roughly by the number of stars that lie an angular distance 
$\Delta\theta$ from one side of a loss ellipse,
divided by $\Delta t$.

This is not quite correct however, since a star may precess
past the loss ellipse before it has had time to reach periapse.
We carry out a more exact calculation 
by assuming that the torus is uniformly populated at some initial time,
with unit total number of stars.
To simplify the calculation, we transform to a new
phase plane defined by
\begin{eqnarray}
\psi &=& \frac{\nu_{x0}\ell_{x0}^2\theta_1 + \nu_{y0}\ell_{y0}^2\theta_2}
{\sqrt{\nu_{x0}^2\ell_{x0}^2 + \nu_{y0}^2\ell_{y0}^2}}, \\
\vartheta &=& \frac{-\nu_{y0}\ell_{x0}\ell_{y0}\theta_1 + \nu_{x0}\ell_{x0}\ell_{y0}\theta_2}
{\sqrt{\nu_{x0}^2\ell_{x0}^2 + \nu_{y0}^2\ell_{y0}^2}}.
\end{eqnarray}
With this transformation, the phase velocity becomes
\begin{equation}
\dot\psi = (\nu_{x0}^2\ell_{x0}^2 + \nu_{y0}^2\ell_{y0}^2)^{1/2},
\ \ \ \ \ \ 
\dot\vartheta = 0
\end{equation}
and the loss regions become circles of radius $\ell_\bullet$.
The angular displacement in one radial period is
\begin{equation}
\Delta\psi = 2\pi(\nu_p/\nu_r)\left(\nu_{x0}^2\ell_{x0}^2 +
\nu_{y0}^2\ell_{y0}^2\right)^{1/2}.
\end{equation}
The density of stars is
$(4\pi^2\ell_{x0}\ell_{y0})^{-1}$. 

At any point in the ($\psi,\vartheta$) plane, stars have a range
of radial phases.
Assuming that the initial distribution satisfies Jeans' theorem,
stars far from the loss regions are  uniformly distributed in $\chi$ where
\begin{equation}
\chi = P^{-1} \int_{r_p}^r \frac{dr}{v_r};
\end{equation}
here $P\equiv 2\pi/\nu_r$ is the radial period, $r_p$ is the periapse distance and $v_r$ is the radial velocity.
The integral is performed along the orbit, hence $\chi$ ranges
between 0 and 1 as $r$ varies from $r_p$ to apoapse and back to
$r_p$. ($\chi = w \mbox{ mod } 2\pi$, where $w$ is mean anomaly).

\begin{figure}[t]
\includegraphics[width=0.5\textwidth]{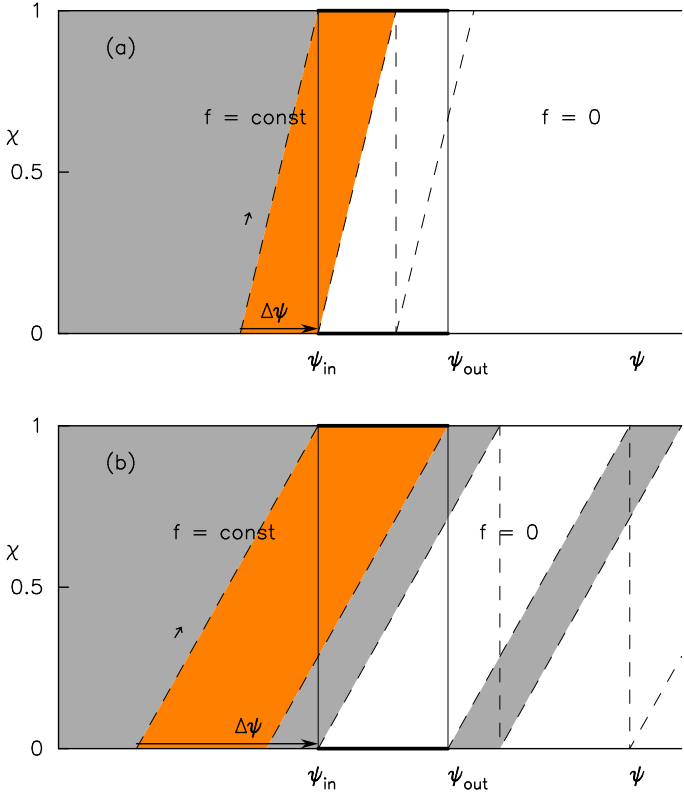}
\caption{Trajectories of stars in the $(\psi,\chi)$ plane as
they encounter a loss region from left to right, 
defined as $\psi_\mathrm{in}\le\psi\le \psi_\mathrm{out}$.
$\chi$ increases from 0 at periapse, to $1/2$ at apoapse, to
$1$ at subsequent periapse.
Trajectories are indicated by dashed lines.
Stars are lost if they reach periapse while inside the loss region.
Stars within the orange region are lost in one radial period.
(a) $\Delta\psi < \psi_\mathrm{in}-\psi_\mathrm{out}$;
(b) $\Delta\psi > \psi_\mathrm{in}-\psi_\mathrm{out}$.
} \label{fig:phase3}
\end{figure}

Figure~\ref{fig:phase3} shows how stars move in the $(\chi,\psi)$ plane at
fixed $\vartheta$.
The loss region extends in $\psi$ a distance
$2\sqrt{\ell_\bullet^2 - \vartheta^2}$, from
$\psi_\mathrm{in}$ to $\psi_\mathrm{out}$.
Stars are lost to the BH if they reach periapse while in
this region.

Two regimes must be considered, depending on whether $\Delta \psi$
is less than or greater than $\psi_\mathrm{out}-\psi_\mathrm{in}$.

1. $\Delta\psi < \psi_\mathrm{out}-\psi_\mathrm{in}$
(Figure~\ref{fig:phase3}a).
In one radial period, stars in the orange region are lost.
One-half of this region lies {\it within} the loss ellipse; these
are stars with $\ell<\ell_0$ but which have not yet attained
periapse.
The persistence of stars inside the ``loss cone'' is similar
to what occurs in the case of diffusional loss cone repopulation,
where there is also a ``boundary layer'', the width of which depends
on the ratio of the relaxation time to the radial period
(e.g.~\cite{CohnKulsrud1978}).
The other one-half consists of stars that have not yet entered
the loss region.
The area of the orange region is equal to the area of a
rectangle of unit height and width $\Delta\psi$;
since stars are distributed uniformly on the $(\chi,\psi)$ plane,
the number of stars lost per radial period is equal to the 
total number of stars, of any radial phase, 
contained within $\Delta\psi$.

2. $\Delta\psi > \psi_\mathrm{out}-\psi_\mathrm{in}$
(Figure~\ref{fig:phase3}b).
In this case, some stars manage to cross the loss region
without being captured.
The area of the orange region is equal to that of a rectangle
of unit height and width $\psi_\mathrm{out}-\psi_\mathrm{in}$.
The number of stars lost per radial period is 
therefore equal to the number of stars, of arbitrary radial phase, 
contained within $\psi_\mathrm{out}-\psi_\mathrm{in} =
2\sqrt{\ell_\bullet^2 - \vartheta^2}$.

To compute the total loss rate, we integrate the loss per
radial period over $\vartheta$.
It is convenient to express the results in terms of $q$ where 
\begin{equation}  \label{eq:defq}
q \equiv \frac{\Delta\psi}{2\ell_\bullet} =
\pi \frac{\nu_p}{\nu_r} \ell_\bullet^{-1}
\sqrt{\nu_{x0}^2\ell_{x0}^2 + 
      \nu_{y0}^2\ell_{y0}^2}.
\end{equation}
$q\ll 1$ corresponds to an ``empty loss cone'' and $q\gg 1$ to a 
``full loss cone''. 
However we note that -- for any $q<1$ -- there are values of
$\vartheta$ such that the width of the loss region,
$\psi_\mathrm{out}-\psi_\mathrm{in}$, is less than $\Delta\psi$.
In terms of the integral $W$ defined above (\ref{UWlat}),
$q$ becomes simply
\begin{equation}  \label{eq:defqH}
q = \frac{P\nu_p}{6\ell_\bullet} \sqrt{W}.
\end{equation}
Unlike the case of collisional loss cone
refilling, where $q=q(E)$ is only a function of energy,
here $q$ is also a function of a second integral $W$.
Pyramid orbits with small opening angles will have small $W$
and small $q$.

The area on the $(\psi,\vartheta)$ plane that is lost, in
one radial period, into one of the four loss regions is
\begin{equation}
2\int_0^{\vartheta_c} \Delta\psi d\vartheta + 
2\int_{\vartheta_c}^{\ell_\bullet} (\psi_\mathrm{out}-\psi_\mathrm{in})d\vartheta
\end{equation}
where
\begin{equation}
\vartheta_c \equiv \ell_\bullet\sqrt{1-q^2}
\end{equation}
is the value of $\vartheta$ where $\Delta\psi = 
\psi_\mathrm{out}-\psi_\mathrm{in}$; 
for $q\ge 1$, $\vartheta_c=0$.
For $q\le 1$, the area integral becomes
\begin{eqnarray}
&&4q\ell_\bullet\int_0^{\vartheta_c} d\vartheta + 
4\int_{\vartheta_c}^{\ell_\bullet} \sqrt{\ell_\bullet^2 - \vartheta^2} d\vartheta \nonumber\\
&=& 4q\ell_\bullet^2\sqrt{1-q^2}\ + 
4\ell_\bullet^2\int_{\sqrt{1-q^2}}^1 dx\sqrt{1-x^2} \nonumber\\
&=& \ell_\bullet^2 \left(\pi + 2q\sqrt{1-q^2} - 2\arcsin\sqrt{1-q^2}\right) \nonumber\\
&=& 4q\ell_\bullet^2 f(q),\nonumber\\
f(q) &=& \frac{1}{2}\sqrt{1-q^2} + \frac{1}{2q}\arcsin(q)
\label{eq:deffofq}
\end{eqnarray}
and for $q>1$ it is $\pi\ell_\bullet^2$.
The function $f(q)$ varies from $f(0)=1$ to $f(1)=\pi/4\approx 0.785$.

\begin{figure}[t]
\includegraphics{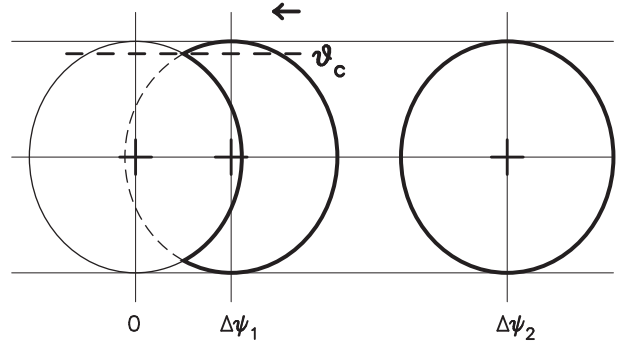}
\caption{Illustrating the area of the $(\psi,\vartheta)$
phase plane that is lost into the BH each radial period.
The circle centered at $(0,0)$ is the loss region corresponding
to one corner of the pyramid orbit;
its radius is $\ell_\bullet$.
Regions marked in bold denote the area of the torus that
is lost in one radial period, for $q<1$ ($\Delta\psi_1$)
and $q>1$ ($\Delta\psi_2$).
While the number of stars lost per radial period is proportional
to the marked areas, the region on the torus from which those
stars come is more complicated since it depends also on an
orbit's radial phase (Figure~\ref{fig:phase3}).
} \label{fig:newphase}
\end{figure}

The area on the phase plane that is lost each radial period 
can be interpreted in a very simple way geometrically, 
as shown in Figure~\ref{fig:newphase}.

Considering that there are four loss regions, the
instantaneous total loss rate $\cal F$, in dimensionless units, is
\begin{subequations}  \label{eq:rate}
\begin{eqnarray}
{\cal F} &=& f(q)\frac{2\ell_\bullet}{\pi^2\,\ell_{x0} \ell_{y0}}
\sqrt{\nu_{x0}^2\,\ell_{x0}^2 + \nu_{y0}^2\,\ell_{y0}^2} 
= \frac{\mu}{P\nu_p} \frac{4q\,f(q)}{\pi} \nonumber\\
&&\mbox{for }0\le q\le 1, \label{eq:rate1} \\
{\cal F} &=& q^{-1}\frac{\ell_\bullet}{2\pi \,\ell_{x0} \ell_{y0}}
\sqrt{\nu_{x0}^2\,\ell_{x0}^2 + \nu_{y0}^2\,\ell_{y0}^2} 
= \frac{1}{2\pi^2} \frac{\ell_\bullet^2}{\ell_{x0}\ell_{y0}} \frac{\nu_r}{\nu_p} \nonumber\\
&=& \frac{\mu}{P\nu_p}
\ \ \ \ \mbox{for }q>1.  \label{eq:rate2}
\end{eqnarray}
\end{subequations}
The second expression for the loss rate, equation~(\ref{eq:rate2}),
can be called the ``full-loss-cone'' loss rate, since it
corresponds to completely filling and empyting the
loss regions in each radial step (Figure~\ref{fig:newphase}).
Note that the loss rate for $q<1$
is $\sim q$ times the full-loss-cone loss rate.
A similar relation holds in the case of collisionally repopulated
loss cones \citep{CohnKulsrud1978}.

The inverse of the loss rate $\cal F$ gives an estimate of the time
 $t_\mathrm{drain}$ required to drain an orbit, or equivalently
the time for a single star, of unknown initial phase, to go into the BH.
In this approximation, the loss rate remains constant until
$t=t_\mathrm{drain}$ at which time the torus is completely empty.
In reality the draining time will always be longer than this,
since after $\sim 1$ precessional periods,
some parts of the torus that are entering the loss regions 
will be empty and the loss rate will drop below equation~(\ref{eq:rate}).
For $\Delta\psi\ge\psi_\mathrm{in}-\psi_\mathrm{out}$, 
the downstream density in Figure~\ref{fig:phase3}, integrated
over radial phase, is easily shown to be
$1-q^{-1}\sqrt{1-\vartheta^2/\ell_\bullet^2}$
times the upstream density while 
for $\Delta\psi<\psi_\mathrm{in}-\psi_\mathrm{out}$
the downstream density is zero.
Integrated over $\vartheta$, the downstream depletion factor becomes
\begin{equation}
1-\frac{\pi}{4q} - \sqrt{1-q^2}(1+q) + \frac{1}{2q}\sin^{-1}\sqrt{1-q^2}
\end{equation}
for $q\le 1$ and $1-\pi/4q$ for $q>1$; it is $0$ for $q=0$,
$\sim 0.215$ for $q=1$ and $1$ for $q\rightarrow\infty$.
For small $q$, the torus will become striated,
containing strips of nearly-zero density interlaced with 
undepleted regions; the loss rate will exhibit discontinous
jumps whenever a depleted region encounters a new loss ellipse
and the time to totally empty the torus will depend in a complicated
way on the frequency ratio $\nu_x/\nu_y$ and on $\ell_\bullet$.
For large $q$, the loss rate will drop more smoothly with time,
roughly as an exponential law with time constant
$\sim t_\mathrm{drain}$.\footnote{This was the approximation
adopted by Merritt \& Poon (2004).}

We postpone a more complete discussion of loss cones in the
triaxial geometry to a future paper.
Here we make a few remarks about pyramids 
with arbitrary opening angles, i.e. for which $e_{x0}, e_{y0}$
are not required to be small.

For each orbit one can compute $\mu$, the fraction of the torus 
occupied by the loss cone (equation~\ref{eq:fraction}), 
by numerically integrating the equations of motion (\ref{eq:Averaged})
and analyzing the probability distribution for instantaneous values of 
$\ell^2$: ${\cal P}(\ell^2 < X) \propto X-\ell_{min}^2$,
where $\ell_{min}^2$ allows for a nonzero lower bound on $\ell^2$. 
Almost all pyramids have $\ell_{min}=0$, but some of them happen to 
be resonances 
(commensurable $\nu_x$ and $\nu_y$) and hence avoid approaching $\ell=0$.
This linear character of the distribution of $\ell^2$ near its minimum 
corresponds to a linear probability distribution of periapse radii 
(${\cal P}(r_\mathrm{peri}<r) \propto r$), 
which is natural to expect if we combine a quadratic distribution 
of impact parameters at infinity with gravitational focusing 
(see equation~7 of \citet{MerrittPoon2004}).

\begin{figure}[t] 
\includegraphics{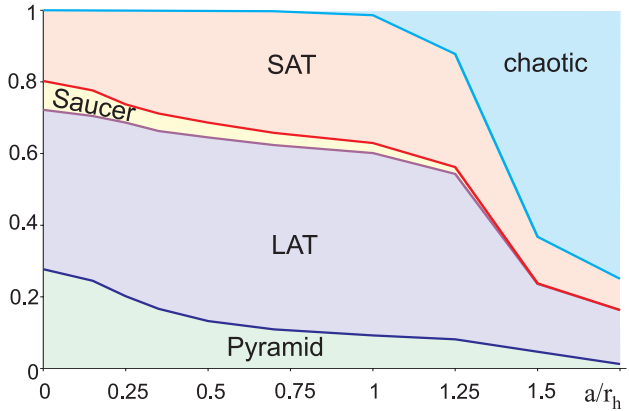}
\caption{Proportion 
of phase space ($\ell^2<\ell_\mathrm{max}^2(0)$) 
that is occupied by the major orbit families:
LATs, pyramids, SATs, saucers, and chaotic orbits,
as a function of semimajor axis $a$ 
based on real-space integrations.
Triaxiality parameters are $\epsilon_b=0.5\epsilon_c$ and 
$\epsilon_c = 0.12/[0.2+(a/\rh)^{-1}]$ (equation~\ref{eq:eps_tilde}), 
corresponding to a density cusp with $\gamma=1$ and 
$\epsilon_c=0.1$ at $a=\rh$.
For $a \gtrsim \rh$ most low-$\ell$ orbits are chaotic
\citep{PoonMerritt2001}.
} \label{fig:proportion_r}
\end{figure}

The coefficient $\mu$ for each orbit is calculated as 
${\cal P}(\ell^2 < \ell_\bullet^2)$.
As seen from equation~(\ref{eq:fraction}), 
the smaller the extent of a pyramid in any direction, 
the greater $\mu$ -- this is true even for orbits with large 
$e_{x0}$ or $e_{y0}$.
While $\mu$ varies greatly from orbit to orbit, its overall distribution 
over the entire ensemble of pyramid orbits follows a power law:
\begin{equation}  \label{mudistr}
{\cal P}_\mu(\mu>Y) \approx (Y/\mu_{min})^{-2} \;, \qquad
\mu_{min} \approx \frac{\ell_\bullet^2}{2\tilde\eta};
\end{equation}
${\cal P}_\mu$ is the probability of having $\mu$ greater than a 
certain value and
$\tilde\eta$ is the fraction of pyramids among all orbits (\ref{eta_pyr}).
The average $\mu$ for all pyramid orbits is therefore 
$\overline{\mu} = 2\mu_{min}$, 
and the average fraction of time that a random orbit of any type and any $\ell$
spends inside the loss cone is 
$\overline\mu \tilde\eta \simeq \ell_\bullet^2$
(almost independent of the potential parameters $\epsilon_b$ and $\epsilon_c$) --
the same number that would result from an isotropic distribution of orbits 
in a spherically-symmetric potential.

\section{Comparison with real-space integrations}  \label{sec_realspace}

We tested the applicability of the orbit-averaged approach by comparing 
the orbit-averaged equations of motion, equation~(\ref{eq:MotionAverage}),
with real-space integrations of orbits having the same initial conditions 
(and arbitrary radial phases).
The agreement was found to be fairly good for orbits with semimajor axes 
$a \lesssim 0.1\rh$: 
about 90\% of the orbits were found to belong to the same orbital class, 
and the correspondence between values of $\mu$ and $\ell_\mathrm{min}^2$ 
was also quite good for individual orbits. 
Averaged over the ensemble, the proportion of phase space occupied by 
the different orbital families,
as well as the net flux of pyramids into the BH, 
is almost the same for the two methods. 
However, at larger radii, the relative fraction of pyramids and 
saucers decreases (Figure~\ref{fig:proportion} (right), 
\ref{fig:proportion_r}). 
Since the maximum possible angular momentum for orbits with a
given semimajor axis $a$ grows faster than $\sqrt{G\mh a}$,
this means that the fraction of pyramid orbits 
among {\sl all} (not just low-$\ell$) orbits is even smaller. 
For orbits with semimajor axis $a\gtrsim \rh$ the frequency of 
radial oscillation becomes comparable to the 
frequencies of precession, and when these overlap, 
orbits tend to become chaotic.
(Weakly chaotic behavior starts earlier). 
So low-$\ell$ orbits with $a > 1.5\rh$ are mostly chaotic, 
as seen from Figure~\ref{fig:proportion_r}, 
confirming that regular pyramid orbits (along with saucers) 
exist only within BH sphere of influence \citep{PoonMerritt2001}.
\footnote{We note that saucer orbits also exist in potentials with high central concentration of mass, 
such as logarithmic potential studied in e.g. \cite{LeesSchwarzschild1992}.}

\section{Effects of general relativity}  \label{sec_GR}

In the previous sections we considered the BH as a Newtonian point mass.
In general relativity (GR), the gravitational field of the 
BH is more complicated, and this will affect the behavior of orbits 
with distances of closest approach that are comparable to $r_g\equiv
GM_\bullet/c^2$.

For a non-spinning BH, the lowest order post-Newtonian effect is advance
of the periapse, which acts in the opposite sense to the precession due 
to an extended mass distribution.
The GR periapse advance is
\begin{equation}
\Delta\varpi = \frac{6\pi}{c^2} \frac{G\mh}{(1-e^2)a}
\end{equation}
per radial period, with $c$ the speed of light
\citep{Weinberg1972},
making the orbit-averaged precession frequency
\begin{equation}  \label{nu_GR}
\nu_{GR} = \nu_r \frac{3 G M_\bullet}{c^2\,a\,\ell^2}.
\end{equation}
We can approximate the effects of this precession by adding an
extra term to the orbit-averaged Hamiltonian (\ref{eq:Averaged}):
\begin{subequations}
\begin{eqnarray}  \label{eq:AveragedGR}
H &=& -\frac{3}{2}\ell^2 + \epsilon_b H_b + \epsilon_c H_c - \frac{\varkappa}{\ell} \\ 
\varkappa &\equiv& \frac{\nu_{GR} \ell^2}{\nu_p}
= \frac{3G\mh}{c^2a}\frac{\nu_r}{\nu_p}
\sim \frac{r_\mathrm{Schw}}{a}\frac{M_\bullet}{M(a)}.  \label{eq:defkappa}
\end{eqnarray}
\end{subequations}
This is equivalent to adding the term $\varkappa/l^2$ to the 
equation of motion for $\varpi$, i.e. to the right hand side of
$d\varpi/d\tau = \partial H/\partial \ell$.
When $\ell=\ell_\mathrm{crit}$, where
\begin{equation}\label{eq:deflcrit}
\ell_\mathrm{crit}=\left(\frac{\varkappa}{3}\right)^{1/3},
\end{equation}
the precession due to GR exactly cancels the precession
due to the spherical component of the distributed mass.
Since the angular momentum of a pyramid orbit
approaches arbitrarily close to zero in the absence of GR, 
there will always come a time when its precession
is dominated by the effects of GR, no matter how small
the value of the dimensionless coefficient $\varkappa$.

We again restrict consideration to the simplified Hamiltonian 
(\ref{eq:HApprox}),
valid for $\epsilon_b,c \ll 1$, $\ell^2 \lesssim \epsilon_c$,
now with the added term due to GR.
This Hamiltonian may be rewritten as 
\begin{eqnarray}  
\frac{5}{2}\epsilon_c - H &=& \left[ \frac{3}{2}\ell^2 + \frac{\varkappa}{\ell} \right] +
\left[ \frac{5}{2}\epsilon_c e_x^2 + \frac{5}{2}(\epsilon_c-\epsilon_b) e_y^2 \right]   \nonumber\\
&\equiv& P(\ell) + Q(e_x,e_y) \,,
\label{HApprox_GR}
\end{eqnarray}
where $P$ and $Q$ denote the expressions in the first and second sets of
square brackets.
The minimum of $P(\ell)$ occurs at $\ell=\ell_\mathrm{crit}$:
\begin{equation}  \label{Pmin}
P_\mathrm{min} = \sqrt[3]{81\varkappa^2/8}.
\end{equation}
The function $Q$ can vary from 0 to some maximum value $Q_\mathrm{max}$ 
due to the limitation that $e_x^2+e_y^2 \le 1$.

Two differences from the Newtonian case are apparent.

\noindent 1) For each value of $(e_x, e_y)$ (and therefore $Q$),
there are now two allowed values of $\ell$.
One of these is smaller than $\ell_\mathrm{crit}$ while the other is greater
(Figure~\ref{fig:Pl}).

\noindent 2) Both the minimum and maximum values of $\ell$ -- both of which
correspond to the maximum value of $P$ (Figure~\ref{fig:Pl}) --
are attained when $Q=0$, i.e. when $e_x=e_y=0$. 
The maximum of $Q$ corresponds to $\ell=\ell_\mathrm{crit}$. 
In the Newtonian case, the minimum of $\ell$ corresponds to the maximum of $Q$.

\subsection{Planar orbits}
\label{sec_planar_GR}
We first consider orbits confined to the $y-z$ plane 
($e_x=0$, hence $\Omega=\pi/2, \ell_z=0$ throughout the evolution). 
Namely, we start an orbit from $\varpi=\pi/2$ ($e_y=0$) and $\ell=\ell_0$.
In the absence of GR, such an orbit would be a LAT for 
$\ell_0 > \frac{5}{3}\epsilon_b$ and a pyramid otherwise.

\begin{figure}[t]
\includegraphics{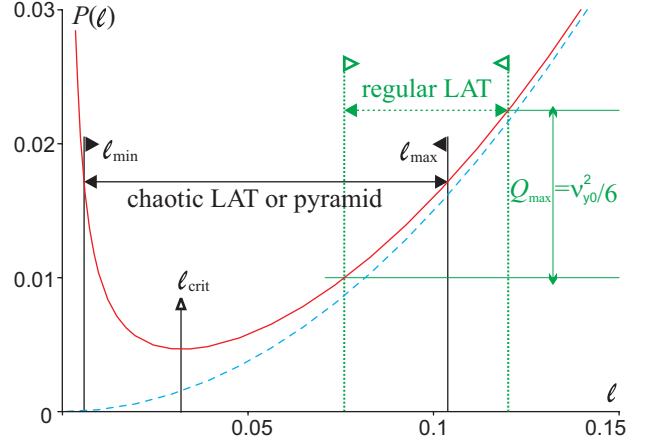}
\caption{
Illustrating the allowed variations in angular momentum $\ell$
for orbits in the presence of general relativistic precession.
The solid (red) curve represents the function $P(\ell)$ 
(equation~\ref{HApprox_GR}); 
the dashed (blue) parabola is the same function in the Newtonian case 
($\varkappa=0$).
If $\varkappa \ne 0$, $P(\ell)$ has a minimum at $\ell_\mathrm{crit}$ 
(equation~\ref{eq:deflcrit}).
Orbits make excursions along the curve $P(\ell)$ in the range from a 
certain value $P_\mathrm{max}$ to $P_\mathrm{max}-Q_\mathrm{max}$ 
(here $Q_\mathrm{max}$ is given for the case of planar orbits of 
\S~\ref{sec_planar_GR} and equals $\frac{5}{2}(\epsilon_c-\epsilon_b) \equiv \nu_{y0}^2/6$). 
If during such an excursion $\ell$ does not cross $\ell_\mathrm{crit}$, 
then the orbit resides on one branch of $P(\ell)$,
typically remaining regular.
Otherwise it flips to the other branch, reaching
lower values of $\ell_\mathrm{min}$ 
(equation~\ref{ell_min}), becoming a (typically chaotic)
LAT or pyramid orbit.
\label{fig:Pl}}
\end{figure}

The Hamiltonian and the equations of motion are 
\begin{eqnarray}  \label{eq:planarGR}
\frac{5}{2}\epsilon_c-H &=& \frac{3}{2}\ell_0^2+\frac{\varkappa}{\ell_0} 
= \frac{3}{2}\ell^2+\frac{\varkappa}{\ell} + \frac{\nu_{y0}^2}{6}\cos^2\varpi \,,\\
\dot \ell &=& -\frac{\nu_{y0}^2}{6} \sin 2\varpi \,,\qquad
\dot \varpi = -3\ell + \frac{\varkappa}{\ell^2}  \,.
\label{eq:planarGReqmotion}
\end{eqnarray}
The orbit in the course of its evolution may or may not attain 
$\varpi=0\ (\mathrm{mod}\,\pi)$.
If it does, then the angle $\varpi$ circulates monotonically, 
with $\dot \varpi \ne 0$.
In Figure~\ref{fig:Pl}, the condition $\dot\varpi=0$ 
corresponds to reaching the lowest point in the $P(\ell)$ curve, 
$\ell=\ell_\mathrm{crit}$. 
Whether this happens depends on the value of $\ell_0$: 
since the orbit starts from $Q=0$ and $P=P(\ell_0)$, 
it can ``descend'' the $P(\ell)$ curve at most by 
$Q_\mathrm{max}=\nu_{y0}^2/6$. 
If this condition is consistent with reaching $P(\ell_\mathrm{crit})$, 
the orbit will flip to the other branch of the $P(\ell)$ curve. 
The condition for this to happen is
\begin{equation}  \label{eq:ell_0pm}
\frac{3}{2}\ell_{0\pm}^2+\frac{\varkappa}{\ell_{0\pm}} = 
\frac{3}{2}\ell_\mathrm{crit}^2+\frac{\varkappa}{\ell_\mathrm{crit}} + \frac{\nu_{y0}^2}{6} ;
\end{equation}
$\ell_{0+}$ and $\ell_{0-}$ are the upper and lower positive roots of this equation. 

If $\ell_0 > \ell_{0+}$, the orbit behaves like a Newtonian LAT 
(Figure~\ref{fig:orbitsgr}, case $a$): 
it has $\dot\varpi < 0$ and $\ell > \ell_\mathrm{crit}$. If $\ell_0 < \ell_{0-}$, the orbit is again a LAT, but now it precesses in the opposite direction 
($\dot\varpi>0$) due to the dominance of GR, 
and $\ell$ never climbs above $\ell_\mathrm{crit}$ 
(Figure~\ref{fig:orbitsgr}, case $e$).
In these cases the condition $\varpi=0$ gives the extremum of $\ell$, 
which is found from equation~(\ref{eq:planarGR}):
\begin{equation}  \label{eq:ell_extrL}
\frac{3}{2}\ell_\mathrm{extr,L}^2+\frac{\varkappa}{\ell_\mathrm{extr,L}} = 
\frac{3}{2}\ell_0^2+\frac{\varkappa}{\ell_0} - \frac{\nu_{y0}^2}{6}  \,.
\end{equation}
This extremum appears to be a minimum ($\ell_\mathrm{extr,L}<\ell_0$) 
if $\ell_0>\ell_{0+}$ and a maximum ($\ell_\mathrm{extr,L}>\ell_0$) 
if $\ell_0<\ell_{0-}$. 

Pyramid orbits are those that reach $\ell=\ell_\mathrm{crit}$.
$\dot\varpi$ changes sign exactly at $\ell_\mathrm{crit}$, but
the angular momentum continues to decrease beyond the point
of turnaround, reaching its minimum value only when $\varpi$
returns again to $\pi/2$, i.e the $z$-axis.
The two semiperiods of oscillation are not 
equal: the first ($\ell>\ell_\mathrm{crit}$ and $\dot\varpi<0$) is slower, 
the other is more abrupt (Figure~\ref{fig:orbitsgr}, cases $b$, $d$).
\footnote{T. Alexander has suggested that these be called
``windshield-wiper orbits.''}
In effect, the orbit is ``reflected'' by striking the GR
angular momentum barrier.
After the orbit precesses past the $z$ axis in the oposite sense,
the angular momentum begins to increase again, reaching its
original value after the precession in $\varpi$ has gone
a full cycle and the orbit has returned to the $z$ axis from the other side.
 
If $\ell_0=\ell_\mathrm{crit}$, there is no oscillation at all -- 
the GR and extended mass precession balance each other exactly
(Figure~\ref{fig:orbitsgr}, case $c$). 
For $\ell_0\lesssim \ell_\mathrm{crit}$ the orbit precesses in the
opposite sense to the Newtonian precession.

We can find the extreme values of $\ell$ by setting $\dot\ell=0$ in
equation~(\ref{eq:planarGReqmotion}).
This occurs for $\varpi=\pi/2$, i.e. for 
$Q=0$ or $P(\ell)=P(\ell_0)$. 
This gives 
\begin{equation}
\ell_\mathrm{extr,P} = \frac{\ell_0}{2} 
\left(\sqrt{1+8\ell_\mathrm{crit}^3/\ell_0^3} - 1\right).
\label{eq:ell_extrP}
\end{equation}
If $\ell_0 > \ell_\mathrm{crit}$, this root corresponds to the minimum $\ell$, 
with $\ell_0$ the maximum value;
in the opposite case they exchange places.
For $\varkappa \ll 3\ell_0^3$ this additional root is
\begin{equation}\label{eq:ell_min_small}
\ell_\mathrm{min} \approx \frac{2\ell_\mathrm{crit}^3}{\ell_0^2} =
\frac{2}{3}\frac{\varkappa}{\ell_0^2}.
\end{equation}
Thus the minimum angular momentum attained by a pyramid orbit
in the presence of GR is approximately proportional to $\varkappa$.
Note the counter-intuitive result that the pyramid orbit with the
widest base (largest $\ell_0$) comes closest to the BH.

Figure~\ref{fig:ell_min_max} shows the dependence of the
maximum and minimum values of $\ell$ on $\ell_0$ for the various
orbit families.

\subsection{Three-dimensional pyramids}

\begin{figure}[t]
\includegraphics{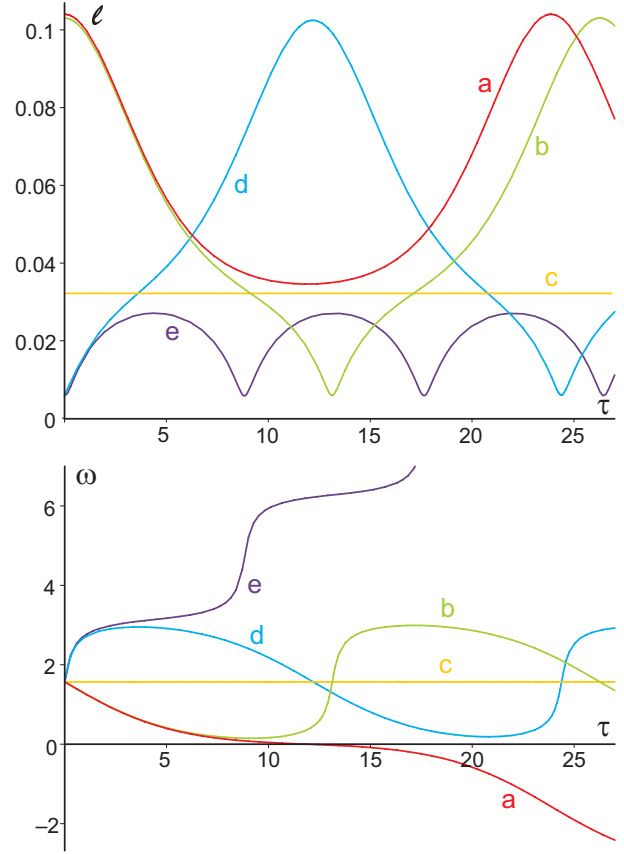}
\caption{
Planar $y-z$ orbits, solutions of equation~(\ref{eq:planarGR}) in a
potential with 
$\epsilon_c=10^{-2}, \epsilon_b=\epsilon_c/2, \varkappa=10^{-4}$, 
started with $\varpi=\pi/2$ and different $\ell_0$:
(a) 0.104, (b) 0.103, (c) 0.0322, (d) 0.006, (e) 0.0058. 
The first two orbits lie close to the separatrix between LATs and pyramids,
$\ell_{0+}=0.10392$ (\ref{eq:ell_0pm}); the
third is the stationary orbit with $\ell_0=\ell_\mathrm{crit}$; and the 
last two lie near the separatrix between pyramids and GR-precession-dominated 
LATs, $\ell_{0-}=0.005845$.
Top panel shows the evolution of $\ell(\tau)$, bottom panel shows $\varpi(\tau)$.
For pyramid orbits (b--d), the angle $\varpi$ librates around $\pi/2$, 
and $\ell$ crosses the critical value $\ell_\mathrm{crit}$; 
tube orbits (a,e) have $\varpi$ monotonically circulating, and $\ell$ is 
always above or below $\ell_\mathrm{crit}$.
\label{fig:orbitsgr}}
\end{figure}

\begin{figure}[t]
\includegraphics{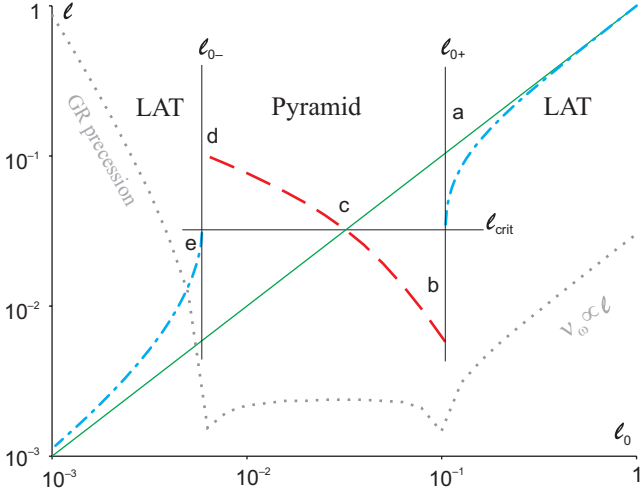}
\caption{
Minimum and maximum values of $\ell$ for a series of orbits with 
initial conditions 
$\ell=\ell_0$, $\omega=\pi/2$, $\ell_z=0$, $\Omega=\pi/2$.
Potential parameters are $\epsilon_c=10^{-2}, \epsilon_b=\epsilon_c/2, \varkappa=10^{-4}$.
The straight line is $\ell=\ell_0$; 
dashed line is the extremum for pyramids, equation~(\ref{eq:ell_extrP}).
These two curves intersect at $\ell_\mathrm{crit}$ 
(equation~\ref{eq:deflcrit}), where they exchange roles. 
For $\ell>\ell_{0+}$ and $\ell<\ell_{0-}$ (equation~\ref{eq:ell_0pm}) 
the orbit is a tube, and the minimum 
(or maximum) is given by equation~(\ref{eq:ell_extrL}).
Dotted grey line shows the leading frequency of $\varpi$ oscillations,
$\nu_\varpi \times 10^{-2}$; 
for high-$\ell$ orbits $\nu_\varpi \approx 3\ell$, 
for orbits dominated by GR precession 
$\nu_\varpi \approx 2\varkappa/\ell^2$.
Letters denote the position of orbits shown in Figure~\ref{fig:orbitsgr}.
\label{fig:ell_min_max}}
\end{figure}

\begin{figure*}[t] 
\includegraphics{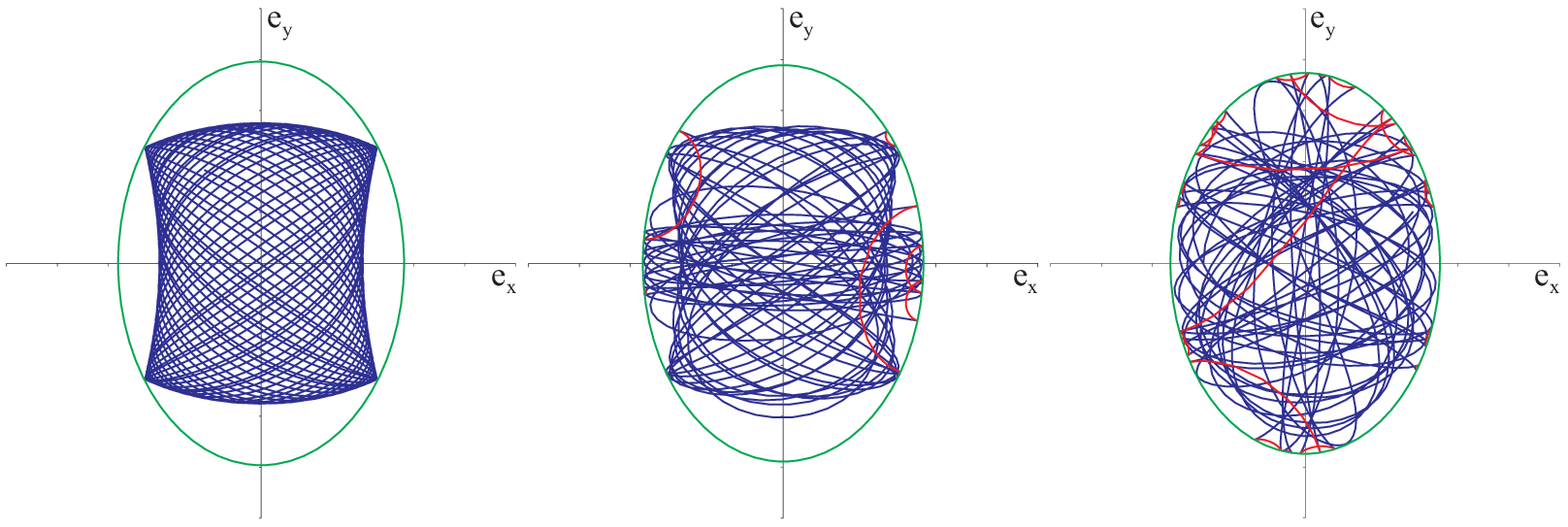}
\caption{Three pyramid orbits with the same initial conditions 
($\ell=0.05, \ell_z=0.02, \varpi=\Omega=\pi/2; 
\epsilon_c=0.01, \epsilon_b=0.005$) 
and three values of the GR coefficient $\varkappa$ 
(equation~\ref{eq:defkappa}).
{\it Left:} $\varkappa=0$ (regular);
{\it middle:} $\varkappa=10^{-6}$ (weakly chaotic); 
{\it right:} $\varkappa=10^{-5}$ (strongly chaotic).
The green ellipse marks the maximal extent of the ($e_x, e_y$) 
vector, equation~(\ref{exey_ellipse}), 
i.e. $\ell=\ell_\mathrm{crit}$, equation~(\ref{eq:deflcrit}); 
red segments correspond to $\ell < \ell_\mathrm{crit}$, 
blue to $\ell > \ell_\mathrm{crit}$ and to the nonrelativistic case. 
} \label{fig:exey_plane}
\end{figure*}

In the case of pyramid orbits that are not restricted to a principal
plane, numerical solution of the equations of motion derived from
the Hamiltonian~(\ref{eq:AveragedGR}) are observed to be generally
chaotic, increasingly so as $\varkappa$ is increased 
(Figure~\ref{fig:exey_plane}).
This may be attributed to the ``scattering'' effect of the GR term 
$\varkappa/l$ in the Hamiltonian, 
which causes the vector $(e_x, e_y)$ to be deflected by 
an almost random angle whenever $\ell$ approaches zero. 
In the limit that the motion is fully chaotic,
$H$ remains the only integral of the motion.
The following argument suggests that the minimum value of the
angular momentum attained in this case should be 
the same as in equation~(\ref{eq:ell_extrP}).

Suppose that the Hamiltonian (\ref{HApprox_GR}) is the only integral that remains.
Then the vector ($e_x, e_y$) can lie anywhere inside an ellipse 
\begin{equation}  \label{exey_ellipse}
Q(e_x, e_y) \equiv \frac{5}{2}\left[ \epsilon_c e_x^2 + (\epsilon_c-\epsilon_b) e_y^2 \right]  \le Q_\mathrm{max} \,,
\end{equation}
whose boundary is given by
\begin{equation}  \label{Qmax}
Q_\mathrm{max} = \frac{5}{2}\epsilon_c - H - P_\mathrm{min}.
\end{equation}
This ellipse defines the base of the ``pyramid'' 
(which now rather resembles a cone).
As in the planar case, the maximum and minimum values of $\ell$ are attained not on the boundary of this ellipse
(i.e. the corners in the Newtonian case), 
but at $e_x=e_y=0$, where $Q=0$ and $P$ attains its maximum.
These values are given by the roots of the equation
$P(\ell) = \frac{5}{2}\epsilon_c - H$, or
\begin{equation}\label{eq:cubic2d}
3\ell^3 - (5\epsilon_c-2H)\ell + 6\ell_c^3=0 \,.
\end{equation}
The two positive roots of this cubic equation are given by
\begin{eqnarray}  \label{ell_min_max_GR}
\ell_\mathrm{min, max} &=& \frac{2}{3}\sqrt{5\epsilon_c-2H}\, \sin\left(\frac{\pi}{6} \pm \phi\right) \;,\\
\phi &=& \frac{1}{3} \arccos \left(\frac{9\varkappa}{(5\epsilon_c-2H)^{3/2}}\right) \;. \nonumber
\end{eqnarray}
The plus sign in the argument of the sine function 
gives $\ell_\mathrm{max}$ while the minus sign gives $\ell_\mathrm{min}$. 
These two values are linked by a simple relation:
\begin{equation}  \label{ell_min}
\ell_\mathrm{min} = \frac{\ell_\mathrm{max}}{2} \left(\sqrt{1+8\varkappa\ell_\mathrm{max}^3/3} - 1\right)
\approx \frac{2}{3} \frac{\varkappa}{\ell_\mathrm{max}^2} \,,
\end{equation}
where the latter approximate equality holds for $\varkappa \ll \ell_\mathrm{max}^3$. In the same approximation
\begin{equation}  \label{ell_min_max_approx}
\ell_\mathrm{min} \approx \frac{2\varkappa}{5\epsilon_c-2H}\;,\quad 
\ell_\mathrm{max}^2 \approx \frac{5\epsilon_c-2H}{3} \,.
\end{equation}
Equation (\ref{eq:ell_extrP}) for planar pyramids is a special case of this relation where $\ell_0=\ell_\mathrm{max}$.

The ellipse (\ref{exey_ellipse}) serves as a ``reflection boundary'' 
for trajectories that come below $\ell \approx\ell_\mathrm{crit}$.
If this happens, the vector ($e_x, e_y$) is observed to be quickly 
``scattered'' by an almost random angle (Figure~\ref{fig:exey_plane}, right, 
denoted by the red segments), similar to the rapid change in 
$\varpi$ that occurs in the planar case (Figure~\ref{fig:orbitsgr}). 
Roughly speaking, all pyramid orbits and some tube orbits 
(those that may attain $\ell \le \ell_\mathrm{crit}$) will be chaotic.
\footnote{A small fraction of the ``flipping'' orbits, 
especially those that oscillate near $\ell_\mathrm{crit}$ 
(close to the lowest point on the $P(\ell)$ curve 
of Figure~\ref{fig:Pl}), 
may retain regularity by virtue of being resonant.}

The distinction between pyramids and chaotic tubes is in the 
radius of this ellipse: 
pyramids by definition have a fixed sign of $e_z$, or $e_x^2+e_y^2<1$, 
which means that 
the ellipse (\ref{exey_ellipse}) should not touch the circle $e_x^2+e_y^2=1$.
Hence pyramids have 
$Q \le \frac{5}{2}(\epsilon_c-\epsilon_b)$, and 
\begin{equation}  \label{H_pyr_GR}
\frac{5}{2}\epsilon_b - P_\mathrm{min} \le H \le \frac{5}{2}\epsilon_c - P_\mathrm{min}.
\end{equation}
This condition is different from the one described in \S\ref{sec_analysis} 
even in the case $\varkappa=0=P_\mathrm{min}$, 
since now pyramids do not coexist with LAT orbits.

The condition for LATs to be chaotic 
(i.e. to pass through $\ell_\mathrm{crit})$ 
is $P_\mathrm{min}+Q_\mathrm{max} \le \frac{5}{2}\epsilon_c-H$.
For LATs the ellipse (\ref{exey_ellipse}) always intersects the unit circle,
so this condition can be satisfied for 
$-P_\mathrm{min} \le H \le \frac{5}{2}\epsilon_b - P_\mathrm{min}$.
However, this is a necessary but not sufficient condition for a chaotic LAT: 
some orbits from this range do not attain $\ell<\ell_\mathrm{crit}$ because of 
the existence of another integral of motion besides $H$ (that is, they are regular).

Finally, we consider the character of the motion
when the precession is dominated by GR, as would be the case
very near the BH. 
This is equivalent to staying on the left branch of $P(\ell)$, with 
$\ell_\mathrm{min,max} \ll \ell_{0-} < \ell_\mathrm{crit}$ (\ref{eq:ell_0pm}).
In this limit there is a second short time scale in addition to the radial period, 
the time for GR precession.
This situation is similar to the high-$\ell$ case (\S~\ref{sec_largel}), 
in the sense that we can carry out a second averaging over $\varpi$ and 
obtain the equations that describe the precession of an annulus due to the triaxial torques.
The orbits in this case are again short- or long-axis tubes.
The only difference from \S~\ref{sec_largel} is that we have to add 
the term $-\varkappa/\ell$ to the averaged Hamiltonian (\ref{eq:Hel}), 
but since $\ell$ is constant in this approximation, the equations of motion 
for $\ell_z, \Omega$ do not change. 
These very-low-$\ell$ regular tube orbits can be easily captured by the BH, 
however their number is very small and we do not consider them when
computing the total capture rate.

We argue in \S 8 that the conservation of $\ell$ for orbits in this limit
can have important consequences for resonant relaxation.

\subsection{Capture of orbits by the BH in the case of GR}

The inclusion of general relativistic precession has the effect
of limiting the maximum eccentricity achievable by a pyramid orbit.
However, if $\ell_\mathrm{min}\le\ell_\bullet$,
an orbit can still come close enough to the BH to be disrupted or captured. 
Introducing the quantity 
$w \equiv \ell_\mathrm{min}/\ell_\bullet$, we can write 
\begin{equation}
w \equiv \frac{\ell_\mathrm{min}}{\ell_\bullet} \simeq \frac{1}{\ell_\bullet} \frac{2\varkappa}{5\epsilon_c-2H}
\gtrsim \frac{3}{5\epsilon_c} \frac{\nu_r}{\nu_p} \frac{\ell_\bullet}{\Theta} \,,
\end{equation}
where we used (\ref{nu_GR}, \ref{eq:defkappa}) 
and set $H=0$ as a lower limit for pyramid orbits
(orbits with the smallest $H$ have the largest 
$\ell_\mathrm{max}$ and the smallest $\ell_\mathrm{min}$). 
Comparison with equation~(\ref{eq:defqH}), 
with $W\le (15\epsilon_c)^2$, shows that
\begin{equation}
w \approx \frac{3\pi}{\Theta}\,q^{-1} .
\end{equation}
Roughly speaking, the condition that stars be captured ($w<1$) 
is equivalent to  
the statement that the loss cone is full ($q>1$). 
This is not a simple coincidence: a full loss cone implies that for
low-$\ell$ orbits the mean change of $\ell$ during one radial period
($\sim \nu_0 \ell_0\,\nu_p/\nu_r$) is of order 
$\ell_\bullet$, while the condition $w=1$
requires that for the lowest allowable $\ell$, 
the GR precession rate (\ref{nu_GR}) 
is comparable to the radial frequency. 
These two conditions are roughly equivalent.

We can express this necessary condition for capture in terms of 
more physically relevant quantities. 
Writing equation~(\ref{eq:nu_p}) as
\begin{equation}
\frac{\nu_p}{\nu_r}\approx \frac{1}{2} \frac{M(a)}{\mh},
\end{equation}
and approximating $\rh$ of equation~(\ref{eq:defrh}) as
\begin{equation}
\rh \approx \frac{G\mh}{\sigma^2} \equiv r_0
\end{equation}
with $\sigma$ the one-dimensional stellar velocity dispersion
at $r=\rh$,
the condition $w\le 1$ becomes
\begin{equation}
\frac{1}{\epsilon_c\sqrt{\Theta}} \frac{\sigma}{c}
\left(\frac{a}{r_0}\right)^{\gamma-7/2} \lesssim 1.
\end{equation}

The Milky Way BH constitutes one extreme of the BH mass distribution.
Writing $\Theta\approx 10$ 
(solar-mass main sequence stars),
$\sigma\approx 10^2$ km s$^{-1}$, and $\gamma=3/2$ gives
\begin{equation}
\epsilon_c^{1/2} \frac{a}{r_0} \gtrsim 10^{-2}
\end{equation}
e.g. for $a=0.1r_0\approx 0.3$ pc, $\epsilon_c\gtrsim 10^{-3}$
is required for stars to be captured.
This is a reasonable degree of triaxiality for the Galactic
center.

At the other mass extreme, we consider the galaxy M87, 
for which $\sigma\approx 350$ km s$^{-1}$ and $\Theta \approx 3$.
Setting $\gamma=0.5$, corresponding to a low-density core,
we find
\begin{equation}
\epsilon_c^{1/3} \frac{a}{r_0} \gtrsim 10^{-1}.
\end{equation}
A dimensionless triaxiality of order unity is reasonable
for a giant elliptical galaxy.

In \S~\ref{sec_estimates} we estimate the loss rate
using the expression (\ref{eq:rate2}) for the full loss cone
rate, with the modification that the fraction of time $\mu$ an 
orbit spends inside the loss cone is now given not by (\ref{eq:fraction}), 
but by 
\begin{equation}  \label{eq:fraction_GR}
\mu \approx \frac{\ell_\bullet^2 - \ell_\mathrm{min}^2}{\ell_0^2 - \ell_\mathrm{min}^2}.
\end{equation}
This relation implies that the instantaneous value of $\ell^2$ is 
distributed uniformly in the range 
$[\ell_\mathrm{min}^2..\ell_\mathrm{max}^2]$. 
This is a good approximation for  chaotic pyramid orbits and a reasonable 
(within a factor of few) approximation for chaotic LATs
(and also for regular orbits).

In the next section we point out the importance of the angular
momentum limit for the rate of gravitational wave events due
to inspiral of compact stellar remnants.


\section{Connection with ``resonant relaxation''}\label{sec_RR}

Resonant relaxation (RR) is a phenomenon that arises in stellar systems 
exhibiting certain regularities in the motion
\citep{RauchTremaine1996, HopmanAlexander2006}.
Due to the discreteness of the stellar distribution, 
torques acting on a test star from all other stars do not cancel exactly, 
and there is a residual torque that produces a change in the angular momentum:
\begin{equation}
\left|\frac{d{\bf L}}{dt}\right| \approx \sqrt{N}\,\frac{Gm}{a} = L_c\,\frac{\sqrt{N}\,m}{M_\bullet} 2\pi P^{-1}
\label{eq:dLdt}
\end{equation} 
(here $m$ is the stellar mass, $P=2\pi/\mu_r$ is the radial period, 
$L_c\equiv \sqrt{GM_\bullet a}$ is the angular momentum of a circular orbit with radius $a$, 
and $N$ is roughly the number of stars within a sphere of radius $a$). 
In a non-resonant system this net torque changes the direction randomly 
after each radial period, 
but in the case of near-Keplerian motion, for example, orbits remain almost the same for many radial periods, 
so the change of angular momentum produced by this torque continues in the same direction for a much longer time, 
the so-called coherence time $t_\mathrm{coh}$, 
until the orientation of either the test star's orbit 
or the other stars' orbits change significantly. 
If this decoherence is due to precession of stars in their mean field, 
then 
\begin{equation}
t_\mathrm{coh}\approx t_\mathrm{M}\equiv\nu_p^{-1} 
\approx \frac{\mh}{m}\frac{P}{N}
\end{equation}
where the relevant precession time is that for an orbit
of average eccentricity.

The total change of ${\bf L}$ during $t_\mathrm{coh}$ is 
\begin{equation}\label{eq:lcoh}
(\Delta L)_\mathrm{coh} \approx \sqrt{N}\, \frac{Gm}{a}\, t_\mathrm{coh} \approx \frac{L_c}{\sqrt{N}} .
\end{equation}
On timescales longer than $t_\mathrm{coh}$ the angular momentum experiences 
a random walk with step size 
$(\Delta L)_\mathrm{coh}$ and time step $t_\mathrm{coh}$.
The relaxation time is defined as the time required for an orbit to change its angular momentum by $L_c$, 
and hence it is given by 
\begin{equation}  \label{eq:trrs}
t_{RR,s} \approx \left(\frac{L_c}{\Delta L}\right)^2 \, t_\mathrm{coh}
 \approx P\,\frac{M_\bullet}{m}.
\end{equation}

The above argument describes ``scalar'' resonant relaxation, 
in which both the magnitude and direction 
of $\mathbf{L}$ can change. 
On longer timescales, precessing orbits fill annuli, 
which also exert mutual torques; 
however, since these torques are perpendicular to $\mathbf{L}$, 
they may change only the direction, 
not the magnitude of $\mathbf{L}$. 
This effect is dubbed vector resonant relaxation (VRR), 
and its coherence time is given by the time required for orbital planes 
to change.
In a spherically symmetric system the only mechanism that changes
orbital planes is the relaxation itself.
\footnote{If the BH is spinning, precession 
due to the Lense-Thirring effect also destroys coherence
\citep{MAMW2010}.}
Hence for VRR the coherence time is given by setting 
$\left|d{\bf L}/dt\right| = L_c/t_\mathrm{coh}$
in equation~(\ref{eq:dLdt}):
\begin{equation}\label{eq:tcohVRR}
t_\mathrm{coh}\equiv t_{\Omega, \mathrm{VRR}} 
\approx \frac{\mh}{m} \frac{P}{\sqrt{N}}
\approx \sqrt{N} t_\mathrm{M}
\end{equation}
and the relaxation time, equation~(\ref{eq:trrs}), becomes
\begin{equation}  \label{eq:trrv}
t_{RR,v} \approx t_{\Omega, \mathrm{VRR}}\approx P\,\frac{M_\bullet}{m\sqrt{N}} 
\end{equation} 
which is $\sim \sqrt{N}$ times shorter 
than the scalar relaxation time.

We begin by comparing RR timescales with timescales for orbital
change due to a triaxial background potential.
Consider a star on a (regular) pyramid orbit confined to the $x-z$ plane. 
It experiences periodic changes of angular momentum $\ell \equiv L/L_c$ 
with frequency $\lesssim \nu_{x0}\nu_p$ (\ref{nu0}) and amplitude 
$\ell_{x0} \lesssim \nu_{x0}/3$ (\ref{eq:ell0}). 
Hence, the typical rate of change of angular momentum is 
\begin{equation}
\frac{dL}{dt} \approx L_c\,\nu_{x0}^2\nu_p/3 \approx L_c\,5\epsilon_c \frac{Nm}{M_\bullet} 2\pi P^{-1} \,.
\end{equation}
Comparison with (\ref{eq:dLdt}) shows that the rate of change of angular 
momentum due to unbalanced torques from the other stars (RR) is greater than 
the rate of regular precession if $\epsilon\sqrt{N} \lesssim 1$.
However, the coherence time for RR is a {\sl typical} precession time of 
stars in the cluster, $\nu_p^{-1}$, 
whereas pyramids change angular momentum on a longer timescale 
$(\nu_p\sqrt{15\epsilon})^{-1}$.
On the other hand, in the case of RR the angular momentum continues to change in a random-walk manner
on timescale longer than $t_\mathrm{coh}$, while in the case of precession in triaxial potential its variation is bounded. 

Next we consider VRR, which corresponds to changes 
in  orbital planes defined by the angles $\Omega$ and $i=\arccos(\ell_z/\ell)$. 
The frequency of orbital plane precession in a triaxial potential, 
$\nu_\Omega$, is $\sim \nu_p\sqrt{\epsilon}$ 
for low-$\ell$ orbits (pyramids and saucers) and even lower for other orbits (Figure~\ref{fig:frequencies}).
The corresponding timescale may be written as 
\begin{equation}\label{eq:tOmega}
t_{\Omega, \mathrm{triax}}\ \gtrsim \frac{\mh}{m} \frac{P}{N\sqrt{\epsilon}}.
\end{equation}
Comparison with the VRR timescale (\ref{eq:trrv}) shows that 
$t_{RR,v}/t_{\Omega,\mathrm{triax}} \lesssim \sqrt{N\epsilon}$.
For the Milky Way, these two timescales are roughly equal at $a\sim 0.5$~pc 
(Figure~\ref{fig:timescales}).
For sufficiently large $N$ the regular precession due to triaxial torques 
goes on faster than the relaxation, 
so the coherence time for VRR is now defined by orbit precession, and the relaxation time itself becomes even longer. 
On the other hand, for small enough $N$ the VRR destroys orientation of orbital planes before they are substantially affected by triaxial torques. 
It seems that VRR in triaxial (or even axisymmetric) systems can be 
suppressed by regular orbit precession; 
we defer the detailed analysis of relaxation for a future study.

So far we have considered the torques arising under RR as
being independent of the torques due to the elongated star cluster.
Suppose instead that we {\it identify} the $\sqrt{N}$ torques that drive RR
with the torques due to the triaxial distortion.
The justification is as follows:
During the coherent RR phase, 
the gravitational potential from $N$ orbit-averaged stars can be
represented in terms of a multipole expansion. 
If the lowest-order nonspherical terms in that expansion happen 
to coincide with the potential generated by a uniform-density 
triaxial cluster, the behavior of orbits in the coherent RR regime would be 
identical to what was derived above for orbits in a triaxial nucleus. 
We stress that this is a contrived model; in general, an expansion of 
the orbit-averaged potential of $N$ stars will contain nonzero dipole, 
octupole etc. terms that depend in some complicated way on radius.
Nevertheless the comparison seems worth making since (as we argue below)
there is one important feature of the motion that should depend only 
weakly on the details of the potential decomposition.

Equating the torques due to RR
\begin{equation}
T_\mathrm{RR}\approx \sqrt{N}\frac{Gm}{r}
\end{equation}
with those due to a triaxial cluster,
\begin{equation}
T_\mathrm{triax} \approx \epsilon \frac{GNm}{r}
\end{equation}
our ansatz becomes
\begin{equation}\label{eq:ansatz}
\epsilon\approx N^{-1/2}.
\end{equation}

As shown above (\S\ref{sec_GR}), 
GR sets a lower limit to the angular  momentum of
a pyramid orbit (equation~\ref{eq:ell_min_small}):
\begin{equation}
\ell_\mathrm{min}\approx \frac{\varkappa}{\ell_0^2}\approx 
\frac{\kappa}{\epsilon} \approx
\frac{r_\mathrm{Schw}}{a} \frac{\mh}{M(a)} \sqrt{N};
\end{equation}
the third term comes from setting 
$\ell_0\approx\ell_\mathrm{max}\approx\sqrt{\epsilon}$,
the maximum value for a pyramid orbit,
while the fourth term uses our ansatz~(\ref{eq:ansatz})
and the definition (\ref{eq:defkappa}) of $\varkappa$.
Expressed in terms of eccentricity,
\begin{equation}\label{eq:emax}
1-e_\mathrm{max}\approx\left(\frac{r_\mathrm{Schw}}{a}\right)^2 
\left(\frac{\mh}{m}\right)^2 \frac{1}{N(a)}.
\end{equation}
There is another way to motivate this result that does not depend on
a detailed knowledge of the behavior of pyramid orbits.
If we require that the
GR precession time:
\begin{equation}
\nu_\mathrm{GR}^{-1} \approx \frac{\ell^2}{\varkappa}\nu_p^{-1}
\end{equation}
(eq.~\ref{nu_GR})
be shorter than the time
\begin{equation}\label{eq:ttorque}
\ell\left|\frac{d\ell}{dt}\right|^{-1} \approx 
\frac{\ell}{\epsilon}\nu_p^{-1}
\end{equation}
 for torques to change $\ell$ by of order itself, then
\begin{equation}\label{eq:fundcrit}
\ell\lesssim \frac{\varkappa}{\epsilon} \approx \sqrt{N}\varkappa
\approx \frac{r_\mathrm{Schw}}{a} \frac{\mh}{M(a)} \sqrt{N}
\end{equation}
as above.
In other words, when $\ell\lesssim \ell_\mathrm{min}$, 
GR precession is so rapid that the $\sqrt{N}$ torques are unable to 
change the angular momentum
significantly over one precessional period.

In order for this limiting angular momentum to be relevant to RR, the timescale
for changes in the background potential should be long compared with 
the time over which an orbit with $\ell\approx \ell_\mathrm{min}$ 
appreciably changes its angular momentum.
As just shown, the latter timescale is 
\begin{equation}
t_\mathrm{GR}\equiv\nu_\mathrm{GR}^{-1}
\approx \frac{\ell^2}{\varkappa}\nu_p^{-1}
\approx \varkappa N \nu_p^{-1}.
\end{equation}
The former timescale is the coherence time for VRR, 
equation~(\ref{eq:tcohVRR}):
\begin{equation}
t_\Omega\approx \frac{\mh}{m}\frac{P}{\sqrt{N}}.
\end{equation}
The condition $t_\Omega\gg t_\mathrm{GR}$ is then
\begin{equation}
\frac{\mh}{m}\frac{P}{\sqrt{N}} \gg 
\varkappa N \nu_p^{-1}
\end{equation}
or
\begin{equation}\label{eq:condd}
\frac{a}{r_\mathrm{Schw}} \sqrt{N} \gg \frac{\mh}{m}.
\end{equation}
Applying this to the center of the Milky Way,
the condition becomes
\begin{equation}
\frac{a}{\mathrm{mpc}}\sqrt{N(a)}\gg 10^2
\end{equation}
which is likely to be satisfied beyond a few mpc
from SgrA$^*$.

On timescales longer than $\sim t_\mathrm{coh}$, 
the torques driving RR will change direction.
This is roughly equivalent in our simple model 
to  changing the orientation of the triaxial ellipsoid,
or to changing $\ell_0$ at fixed $\ell$.
Such changes might induce an orbit to evolve to values of $\ell$
lower than $\ell_\mathrm{min}$, by advancing
down the narrow ``neck'' in the lower left portion of 
Figure~\ref{fig:ell_min_max}.
However such evolution would be disfavored, for two reasons:
(1) it would require a series of correlated changes in the
background potential, increasingly so as $\ell$ became small;
(2) as $\ell$ decreased and 
$\nu_\mathrm{GR}$ increased,
changes in the background potential would  occur on timescales
progressively longer than the GR precession time, and
adiabatic invariance would  tend to preserve $\ell$ (\S4.4).
These predictions can in principle be tested via direct
$N$-body integration of small-$N$ systems including
post-Newtonian accelerations \citep[e.g][]{MAMW2010}.

A lower limit to the angular momentum for orbits near a massive BH
could  have important implications for the rate of gravitational
wave events due to  extreme-mass-ratio inspirals, or EMRIs 
\citep{HilsBender1995}.
The critical eccentricity at which the orbital evolution of a
$10M_\odot$ compact object begins to be dominated by gravitational wave
emission is
\begin{equation}
1-e_\mathrm{EMRI}\approx 10^{-5} 
\left(\frac{t_r}{10^9\mathrm{yr}}\right)^{-2/3}
\left(\frac{\mh}{10^6M_\odot}\right)^{4/3}
\end{equation}
with $t_r$ the  relaxation time 
\citep[e.g.][eq.~(6)]{Amaro2007}.
By comparison, equation~(\ref{eq:emax}), after substitution of
$N(<a)=N_0 (a/\mathrm{mpc})$ implies
\begin{equation}
1-e_\mathrm{max} \approx 2\times 10^{-4} \left(\frac{\mh}{10^6M_\odot}\right)^4
\left(\frac{N_0}{100}\right)^{-1}
\left(\frac{a}{10\mathrm{mpc}}\right)^{-3}.
\end{equation}

\section{Estimates for real galaxies}  \label{sec_estimates}

In this section we estimate 
the fraction and lifetime of pyramid orbits
to be expected in the nuclei of real galaxies.

We restrict calculations to the case of ``maximal triaxiality,''
$\epsilon_b = \epsilon_c/2$, although we leave the amplitudes of
$\epsilon_b, \epsilon_c$ free parameters. 
We also limit the discussion to orbits within the BH influence
radius, $r\lesssim\rh$, where our analysis is valid and where
orbits are typically regular\footnote{Excepting for the effects of GR, 
which as noted above may introduce chaotic behavior even for 
 orbit-averaged parameters. The chaos that sets in at $r\gtrsim \rh$ 
(\S~\ref{sec_realspace}) arises from the coupling of the orbit-averaged 
and radial motions.}.
Beyond $\sim\rh$, centrophilic (mostly chaotic) orbits still
exist and could dominate, e.g., the rate of feeding of a central
BH \citep{MerrittPoon2004}.

The first set of parameters is chosen to describe the
center of the Milky Way.
The BH mass is set to $M_\bullet = 4\times 10^6~M_\odot$ 
\citep{Ghez2008,Gillessen2009a,Gillessen2009b}.
The density of the spherically symmetric stellar cusp is taken to be
$\rho_s = 1.5\times 10^5\,M_\odot \mathrm{pc}^{-3}\,(r/1\mbox{ pc})^{-\gamma}$ \citep{Schoedel2007}, 
with $\gamma=1.5$ \citep{Schoedel2008}; 
the corresponding BH influence radius is $\rh \approx 3$~pc.\footnote{
These cusp parameters correspond to an inward extrapolation of the 
density observed at $r\gtrsim 1$ pc.
Recent observations \citep{Buchholz2009,Do2009,Bartko2010}
reveal a ``hole'' in the density of evolved stars inside $\sim 0.5$ pc,
implying a possibly much lower density for the spherical component
near the BH.}
The triaxial component of the potential is highly uncertain;
one source would be the nuclear bar with density 
$\rho_t \approx 150\,M_\odot \mathrm{pc}^{-3}$ \citep{Rodriguez2008},
yielding a triaxiality coefficient at $r=1$~pc of $\epsilon_c \approx 10^{-3}$. 
We also considered a  larger value, $\epsilon_c=10^{-2}$, 
which may be justified by some kind of asymmetry 
on spatial scales closer to $\rh$ than the bar.
In this model, the precession time due to the spherical component of the 
potential, $2\pi/(3\nu_p)$ for a circular orbit, is
independent of radius and equals 
$\sim 1.7\times 10^5$~yr;
the two-body relaxation time is also constant ($5\times 10^9$~yr), 
and timescales for scalar and vector resonant 
relaxation are given by equation~(\ref{eq:trrs}), (\ref{eq:trrv}) 
with stellar mass $m=1\,M_\odot$.

\begin{figure}[t] 
$$\includegraphics{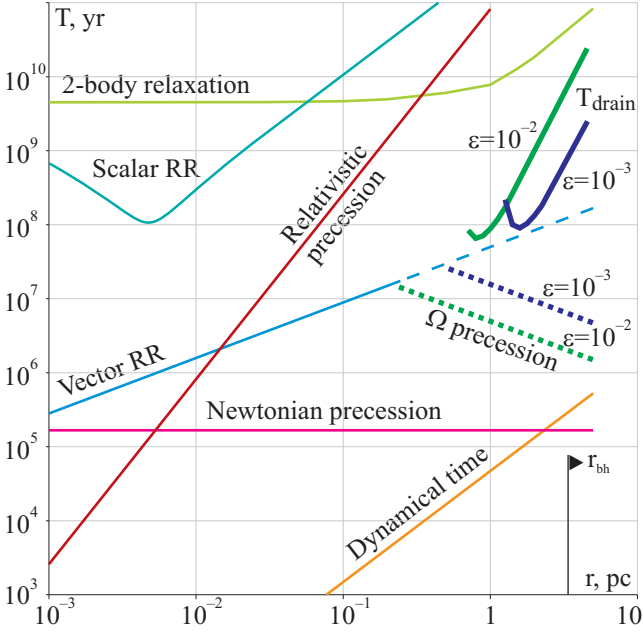}$$
\caption{Various timescales for Milky Way center.
Green and blue solid curves are the pyramid draining times (more exact calculation than equation~\ref{T_dr}) for
$\epsilon_c=10^{-2}$ and $10^{-3}$; dotted green and blue curves denote typical orbital plane precession 
timescales $\nu_\Omega^{-1}$ for low-$\ell$ orbits ($\ell^2\approx \epsilon_c(a)$); 
Newtonian and relativistic precession times are given for circular orbits;
other timescales are marked on the plot.  
Vector resonant relaxation is suppressed when $t_{RR,v} \gtrsim \nu_\Omega^{-1}$ 
(marked by conversion of line into dashed).
} \label{fig:timescales}
\end{figure}

The second set of parameters is intended to describe
the case of galaxies with more massive BHs, 
using the so-called $M_\bullet-\sigma$ relation 
in the form
\begin{equation}  \label{Mbhsigma}
M_\bullet \approx 1.7\times 10^8\,M_\odot\, 
\left(\frac{\sigma}{200\mbox{ km s}^{-1}}\right)^{4.86} 
\end{equation}
\citep{FF2005}.
Combined with the definition of $\rh \approx GM_\bullet/\sigma^2$, we get
\begin{equation}  \label{rbh}
\rh \approx 13\mbox{ pc} \left(\frac{\mh}{10^8\,M_\odot}\right)^{0.59} \;.
\end{equation}
The two-body relaxation time evaluated at $\rh$ 
(assuming a mean-square stellar mass $m_\star=1\,M_\odot$ and a
Coulomb logarithm $\ln\Lambda=15$) is
\begin{subequations}
\begin{eqnarray}  \label{T2br}
t_{2br}(\rh) &\approx& 2.1\times 10^{13}\mbox{ yr} \left(\frac{\sigma}
{200\mbox{ km s}^{-1}}\right)^{7.5} \\
&\approx& 9.6\times 10^{12}\mbox{ yr} \left(\frac{\mh}{10^8\,M_\odot}\right)^{1.54} 
\end{eqnarray}
\end{subequations}
\citep{MMS2007}.

We first estimate the radius $r_\mathrm{crit}$ that separates the empty 
($q<1$) and full loss cone regimes. 
As noted in the previous section, GR precession 
prevents a pyramid orbit from reaching arbitrarily low angular momenta; 
the radius beyond which capture becomes possible 
is roughly $r_\mathrm{crit}$.
Using equation~(\ref{eq:defqH}) with $W=(15\epsilon_c)^2$ 
(the maximum value for pyramids) and 
equations~(\ref{eq:nu_p}), (\ref{ell_bullet}), 
the condition $q=1$ translates to
\begin{equation}  \nonumber
1 = \frac{4\pi}{3-\gamma} \frac{\rho_s r_0^3}{M_\bullet} \left(\frac{r_\mathrm{crit}}{r_0}\right)^{3-\gamma} 
\frac{2\alpha'}{3(2-\gamma)}\, \frac{5\pi\, \epsilon_c(r_\mathrm{crit})}{\sqrt{\Theta r_\mathrm{Schw}/r_\mathrm{crit}}}
\end{equation}
If we take $r_0$ to be $\rh$, and $\rho_s$ and $\sigma$ as the 
density and velocity dispersion at this radius, we obtain
\begin{equation}  \label{r_crit}
\frac{r_\mathrm{crit}}{\rh} \approx 0.5 \left( \frac{\sigma}{c}\frac{\sqrt{\Theta}}{\epsilon_c(r_0)} \right)^{2/7}.
\end{equation}
The radius $r_\mathrm{crit}$ typically lies in the range $(0.2-0.7)\rh$, 
weakly dependent on the parameters.
Since regular pyramid orbits exist only for $a \lesssim \rh$ 
(Figure~\ref{fig:proportion_r}), 
there is evidently a fairly narrow range of radii for which capture of 
stars from 
pyramid orbits is possible.\footnote{
This is also roughly the radial range from which extreme-mass-ratio
inspiral events are believed to originate; \cite[e.g.][]{Ivanov2002}.}
However pyramid-like, centrophilic can exist at much larger radii 
\citep{PoonMerritt2001}.

Next we make a rough estimate of the pyramid draining time at 
$a>r_\mathrm{crit}$, using the expression (\ref{eq:rate2})
for the flux in the full loss cone regime, ${\cal F} = \mu/(P\nu_p)$;
$\mu$ (the fraction of phase space  occupied by the loss cone) 
is given by equation~(\ref{eq:fraction_GR}) with 
$\ell_\mathrm{min} \ll \ell_\bullet$, 
$\ell_\mathrm{max}^2 \approx \frac{5}{3}\epsilon_c$ 
(equation~\ref{ell_min_max_approx}):
\begin{eqnarray}  \label{T_dr}
t_\mathrm{drain} &=& \frac{1}{{\cal F}\,\nu_p} \approx \frac{5\pi}{3\Theta}\frac{c^2}{(GM_\bullet)^{3/2}} a^{5/2} \epsilon_c(a) \\
&\approx& 10^9\mbox{ yr} \times \frac{\epsilon_c(a)}{\Theta} \left(\frac{M_\bullet}{10^8\,M_\odot}\right)^{-3/2} 
\left(\frac{a}{1\mbox{ pc}}\right)^{5/2}.   \nonumber
\end{eqnarray}
A more exact calculation of $t_\mathrm{drain}(a)$ for the Milky Way, 
based on numerical analysis of properties of orbits sampled from the 
entire phase space, is shown in Figure~\ref{fig:timescales}.

Finally, we estimate the total capture rate for all pyramids inside $\rh$, 
using $t_\mathrm{pyr}\equiv t_\mathrm{drain}(\rh)$ as a typical timescale and applying 
(\ref{Mbhsigma}, \ref{rbh}, \ref{T_dr}):
\begin{equation}  \label{Tpyr}
t_\mathrm{pyr} \approx 6\times 10^{11}\mbox{ yr} \times \frac{\epsilon_c(\rh)}{\Theta} 
\left(\frac{M_\bullet}{10^8\,M_\odot}\right)^{-0.025}.
\end{equation}
The capture rate from pyramids is then
\begin{equation}  \label{Mdotpyr}
\dot M_\mathrm{pyr} \approx \frac{\epsilon_c(\rh) M_\bullet}{t_\mathrm{pyr}} 
\approx 1.6\times 10^{-4}\,M_\odot\mbox{ yr}^{-1} \Theta \left(\frac{M_\bullet}{10^8\,M_\odot}\right)^{1.025} 
\end{equation}
For the Milky Way  we find
$\sim 4\times 10^{-5}\,M_\odot$yr$^{-1}$ for $\epsilon_c=10^{-3}$ and 
$\sim 10^{-4}\,M_\odot$yr$^{-1}$ for $\epsilon_c=10^{-2}$.

This capture rate should be compared with that due to two-body relaxation, 
which is estimated to be 
\citep{Merritt2009}
\begin{equation}  \label{Mdot2br}
\dot M_{2br} \approx 0.1\frac{M_\bullet}{t_{2br}} 
\approx 10^{-6}\, M_\odot\mbox{yr}^{-1} \times \left(\frac{M_\bullet}{10^8\,M_\odot}\right)^{-0.54}
\end{equation}
Thus even for a Milky Way-sized galaxy, the capture rate of pyramids could
be  comparable with or greater than that due to two-body relaxation. 
For more massive galaxies this inequality becomes even stronger.
However, this is only the initial capture rate -- after 
$\sim t_\mathrm{pyr}$, all stars on pyramid orbits would have been
consumed, at least in the absence of other mechanisms for 
repopulating the small-$\ell$ parts of phase space
(not necessarily $\ell \lesssim \ell_\bullet$, but the much broader 
region $\ell \lesssim \sqrt{\epsilon_c}$ from which draining is effective). 

In the most luminous galaxies, like M87, standard mechanisms for
relaxation are expected to be ineffective even over Gyr timescales
and pyramid orbits once depleted are likely to stay depleted.
Setting $\epsilon_c=0.1$, $\Theta=3$ and $\mh=4\times 10^9M_\odot$
gives for M87 $t_\mathrm{pyr}\approx 5$ Gyr and $M_\mathrm{pyr}\approx
4\times 10^8M_\odot$.
This could be an effective mechanism for creating a low-density
core at the centers of giant elliptical galaxies.

\section{Conclusions}\label{sec_summary}

We discussed the character of orbits within the radius of influence 
$\rh$ of a supermassive BH at the center of a triaxial star cluster. 
The motion can be described as a perturbation of Keplerian motion;
we derive the orbit-averaged equations and explore
their solutions both analytically (when 
the triaxiality is small) and numerically. 
Orbits are found to be mainly regular in this region.
There exist three families of tube orbits; a fourth orbital family,
the pyramids, can be described as eccentric Keplerian ellipses
that librate in two directions about  the short axis of the triaxial
figure.
At the ``corners'' of the pyramid, the angular momentum reaches
zero, which means that stars on these orbits can be captured by the BH. 
We derive expressions for the rate at which stars on pyramid orbits
would be lost to the BH; there are many similarities with the
more standard case of diffusional loss cone refilling, 
but also some important differences, due  to the fact that 
the approach to the loss cone is deterministic for the pyramids,
rather than statistical.
The inclusion of general relativistic precession is shown to
impose a lower bound on the angular momentum.
We argue that a similar lower bound should  apply to orbital
evolution in the case that the torques are due to resonant
relaxation.
The rate of consumption of stars from pyramid orbits is likely to
be substantially greater than the rate due to two-body relaxation
in the most luminous galaxies,
although in the absence of mechanisms for orbital repopulation,
these high consumption rates would  only be maintained
until such a time as the pyramid orbits have been drained;
however the latter time can be measured in billions of years.

\acknowledgments
We thank T. Alexander, S. Hughes, A. Rasskazov and B. Kocsis for fruitful discussions.
DM was supported by grants AST-0807910 (NSF) and NNX07AH15G (NASA).
EV acknowledges support from Russian Ministry of science and education (grants No.2009-1.1-126-056 and P1336).

\end{document}